\documentclass[11pt,a4paper]{article}
\pdfoutput=1
\usepackage{jheppub}

\usepackage{graphicx}
\usepackage{float}

\usepackage{amssymb}
\usepackage{amsmath}

\subheader{\small OUTP-15-26P \\ DCPT/15/128 \\ IPPP/15/64}

\title{The prompt atmospheric neutrino flux in the light of LHCb}

\author[a]{Rhorry Gauld,}
\author[b]{Juan Rojo,}
\author[b]{Luca~Rottoli,}
\author[b,c]{Subir Sarkar}
\author[b]{and Jim Talbert}

\affiliation[a]{Institute for Particle Physics Phenomenology, 
Durham University,  Durham DH1 3LE, UK}
\affiliation[b]{Rudolf Peierls Centre for Theoretical Physics, 1 Keble Road,  
University of Oxford, OX1 3NP Oxford, UK}
\affiliation[c]{Niels Bohr Institute, Copenhagen University,
Blegdamsvej 17, 2100 Copenhagen, Denmark}

\emailAdd{rhorry.gauld@durham.ac.uk}
\emailAdd{juan.rojo@physics.ox.ac.uk}
\emailAdd{luca.rottoli@physics.ox.ac.uk}
\emailAdd{subir.sarkar@physics.ox.ac.uk}
\emailAdd{jim.talbert@physics.ox.ac.uk}

\abstract{The recent observation of very high energy cosmic neutrinos
  by IceCube heralds the beginning of neutrino astronomy.  At these
  energies, the dominant background to the astrophysical signal is the
  flux of `prompt' neutrinos, arising from the decay of charmed mesons
  produced by cosmic ray collisions in the atmosphere.  In this work
  we provide predictions for the prompt atmospheric neutrino flux in
  the framework of perturbative QCD, using state-of-the-art Monte
  Carlo event generators.  Our calculation includes the constraints
  set by charm production measurements from the LHCb experiment at 7
  TeV, recently validated with the corresponding 13 TeV data.  Our
  result for the prompt flux is a factor of about 2 below the previous
  benchmark calculation, in general agreement with other recent
  estimates, but with an improved estimate of the uncertainty.  This
  alleviates the existing tension between the theoretical prediction
  and IceCube limits, and suggests that a direct detection of the
  prompt flux is imminent.}

\keywords{Neutrino Telescopes, Atmospheric Neutrinos, QCD Phenomenology}

\arxivnumber{1511.06346}

\begin{document}
\maketitle
\flushbottom

\section{Introduction}
\label{sec:introduction}

The IceCube experiment~\cite{Halzen:2010yj} at the South Pole has
recently made the first detection of high energy cosmic neutrinos from
the Southern sky with deposited energies between 30 and 2000 TeV and
arrival directions consistent with
isotropy~\cite{Aartsen:2013bka,Aartsen:2013jdh,Aartsen:2014gkd}. Although
these are mainly $\nu_e$ and $\nu_\tau$ charged-current and
neutral-current (`cascade') neutrino interactions, the 37 events are
consistent with expectations for equal fluxes of all three neutrino
flavours \cite{Aartsen:2015ivb}. Subsequently cosmic $\nu_\mu$
charged-current (`track') events have also been seen from the Northern
sky~\cite{Aartsen:2015rwa} with comparable
flux~\cite{Aartsen:2015knd}. At these high energies, the
`conventional' atmospheric neutrino flux, from the decays of pions and
kaons produced by the collisions of cosmic rays with nuclei in the
atmosphere~\cite{Barr:2004br,GonzalezGarcia:2006ay,Honda:2006qj,Honda:2011nf},
is suppressed due to energy loss before the decays occur. However
charmed mesons decay almost instantaneously and therefore, despite
their smaller production cross-section, the `prompt' neutrino flux
from their decays ($\propto E_{\nu}^{-2.7}$) should dominate over the
conventional flux ($\propto E_{\nu}^{-3.7}$) at high energies. Thus
the prompt component is the most relevant background for the similarly
hard spectrum expected for the astrophysical neutrino flux
\cite{Halzen:2013dva,Anchordoqui:2013dnh}. Indeed the statistical
significance ($5.7\sigma$) with which an atmospheric origin can be
rejected for the 37 IceCube events is limited by the uncertainty in
the expected atmospheric prompt neutrino flux.

Many calculations of the prompt neutrino flux have been
presented~\cite{Lipari:1993hd,Pasquali:1998ji,Enberg:2008te,Gondolo:1995fq,Martin:2003us,
  Gelmini:1999ve,Bhattacharya:2015jpa,Engel:2015dxa,Fedynitch:2015zma,Arguelles:2015wba,Garzelli:2015psa},
however so far IceCube has not detected it and set only an upper limit
of 1.52 times the central value of the benchmark ERS
calculation~\cite{Enberg:2008te} at 90\% CL \cite{Aartsen:2014muf}. In
a recent analysis this limit has been lowered further by a factor of 3
to only 0.54 times the above benchmark calculation
\cite{Schoenen:2015ipa}. This motivates a re-evaluation with
state-of-the-art tools and inputs, providing, in particular, an
improved estimate of all theoretical uncertainties. The major
uncertainty is in the calculation of charm production at high-energies
due to higher-order corrections and, especially, the imprecise
knowledge of the gluon parton distribution function (PDF) at
small-$x_\text{Bjorken}$. A recent breakthrough has been the
availability of charm hadroproduction data from the LHCb
experiment~\cite{Aaij:2013noa,Aaij:2013mga} which covers the
kinematical range directly relevant to the calculation of the
atmospheric prompt neutrino flux.

With this motivation, we have recently validated state-of-the-art
perturbative QCD calculations with the LHCb forward charm production
data at 7 TeV~\cite{Aaij:2013noa,Aaij:2013mga}, and included these
measurements into the NNPDF3.0 global analysis~\cite{Ball:2014uwa}.
We were thus able to construct a new PDF set, NNPDF3.0+LHCb, which is
tailored for calculation of the prompt neutrino flux.\footnote{See
  also~\cite{Zenaiev:2015rfa} for a similar analysis performed in the
  {\tt HERAfitter} framework~\cite{Alekhin:2014irh}.}  We benchmarked
three other codes, {\tt FONLL}~\cite{Cacciari:2001td}, {\tt
  POWHEG}~\cite{Nason:2004rx,Frixione:2007vw,Alioli:2010xd} and {\tt
  aMC@NLO}~\cite{Alwall:2014hca}, finding good agreement both amongst
themselves and with the LHCb data.  Our calculations have subsequently
been found to be in good agreement with the 13 TeV LHCb charm
production data~\cite{Aaij:2015bpa}, which probe even smaller values
of $x$.

In this work we calculate the atmospheric prompt neutrino flux using
the canonical set of cascade equations implemented in the `$Z$-moment'
framework (see~\cite{Enberg:2008te} and references therein). Charm
cross-sections and decays are obtained using the NLO Monte Carlo
generator {\tt POWHEG} with the NNPDF3.0+LHCb PDF set as input.  We
consider several parameterisations of the cosmic ray flux, including
the most recent models~\cite{Gaisser:2013bla,Stanev:2014mla}, and
study the dependence of our result on the choice of input PDF set.

We compare our calculation with previous results, in particular the
benchmark ERS calculation~\cite{Enberg:2008te}, as well as the recent
BERSS~\cite{Bhattacharya:2015jpa} and GMS~\cite{Garzelli:2015psa}
analyses. We emphasise that our calculation is the only one which is
directly validated with LHCb data.  All four calculations are
consistent within our theoretical uncertainty band, with ERS being at
the edge of the upper limit. Our central value is similar to the BERSS
result, while the GMS result is a little higher. We also compare our
result to the recent IceCube limit on the prompt neutrino flux,
finding that our central value is \emph{consistent}. Moreover our
lower limit indicates that the prompt neutrino flux will soon be
detected, thus enabling reliable subtraction of any contamination of
the astrophysical neutrino candidates.

This paper is organised as follows. In \S~\ref{sec:zmomemts} we
discuss the various inputs that enter the calculation of the prompt
neutrino flux, including the parameterisations of the cosmic ray flux,
the solution of the cascade questions, and the calculations of the
various $Z$ moments. The results for the prompt flux are presented in
\S~\ref{sec:results}, where we compare with other recent
determinations as well as with the latest IceCube limit.  We also
study the dependence of our result on the input PDF set and on the
cosmic ray flux parameterisation.  Our results are summarised in
\S~\ref{sec:delivery} and are made publicly available in the form of
an interpolation code which returns the prompt neutrino flux and its
uncertainty for all adopted models of the cosmic ray flux.

\section{Calculation of the prompt neutrino flux}
\label{sec:zmomemts}

First we present the parameterisations of the cosmic ray flux used in
this work. We review the cascade equations for the propagation of
particles in the atmosphere, and their solution using
$Z$-moments. Then we discuss the calculation of the various
$Z$-moments, with emphasis on the charm production cross-section and
the associated theory uncertainties.

\subsection{The incoming cosmic-ray flux}
\label{sec:crflux}

The flux of cosmic rays incident on the atmosphere is dominated by
protons and has been measured by a variety of experiments~(see recent
reviews
\cite{Seo:2012pw,Kampert:2012mx,Gaisser:2013bla,Stanev:2014mla}).
At energies $\gtrsim 10^3$ GeV relevant for calculating the prompt
neutrino flux, a traditional parameterisation has been the broken
Power Law (BPL) with a `knee' at $E_p = 5 \times 10^{6}$ GeV, assuming
all cosmic rays are protons:
\begin{equation}
\label{eq:power}
\phi^{(0)}_p(E) = \begin{cases}
  1.7 E_p^{-2.7} & {\rm GeV}\,{\rm sr}^{-1}\,{\rm cm}^{-2}{\rm s}^{-1},\qquad E_p \le 5\times 10^{6} \,\,{\rm GeV}\\
  174 E_p^{-3} & {\rm GeV}\,{\rm sr}^{-1}\,{\rm cm}^{-2}{\rm s}^{-1}, \qquad  E_p \ge 5\times 10^{6} \,\,{\rm GeV}\\
\end{cases}
\end{equation}
Recently, more elaborate parameterisations of the cosmic ray flux have
been provided, with emphasis on including composition data and
improving the description above the `knee' in the spectrum.  One
such set~\cite{Gaisser:2011cc} follows Hillas'
proposal~\cite{Hillas:2006ms} for accommodating three populations of
cosmic rays: one associated with acceleration by supernova remnant
shocks, a second galactic component from unspecified sources, and
finally an extra-galactic component which dominates at the highest
energies. The assumption is that 5 groups of nuclei, $i$, are
contained in each of these three source components, $j$, such that the
cosmic ray spectrum for the nuclear species $i$ can be written as
\begin{equation}
\phi^{(0)}_{i}(E) = \sum_{j=1}^3 \left[a_{ij}\textit{ } E_i^{-\gamma_{ij}} \exp \left( -\frac{E_i}{Z_{i}R_{c,j}}\right)\right] \, ,
\end{equation}
where $R_{c,j}$ is the magnetic rigidity for the source component $j$,
and $a_{ij}$ and $\gamma_{ij}$ are the corresponding normalization
constants and spectral indices~\cite{Gaisser:2011cc}.

We construct an equivalent `all-proton' spectrum\footnote{We assume
  isospin symmetry, hence `protons' refer to nucleons in general.}
by re-weighting the various nuclei:
\begin{equation}
\phi^{(0)}_{p,i}(E_p)  = A_i \times \phi^{(0)}_{i}(A_i E_{p}) \ ,
\end{equation}
with $A_i$ the atomic number of species $i$, and, to obtain the total
cosmic ray flux, we sum over each of the 5 nuclear components:
\begin{equation}
\phi^{(0)}_{p}(E_p) = \sum_{i=1}^5 \phi^{(0)}_{p,i}(E_i)= \sum_{i=1}^5 \left[
A_i \phi^{(0)}_{i}(A_i E_{p})\right] \ .  
\end{equation}
Thus we do not need to consider an effective nuclear attenuation
length, since collective effects in nucleus-nucleus collisions can be
safely ignored when calculating the nucleon interaction lengths inside
the projectile. (Henceforth we drop the subscript $p$ on $E_p$ except
when essential.)

\begin{figure}[t]
\centering 
\includegraphics[scale=0.4]{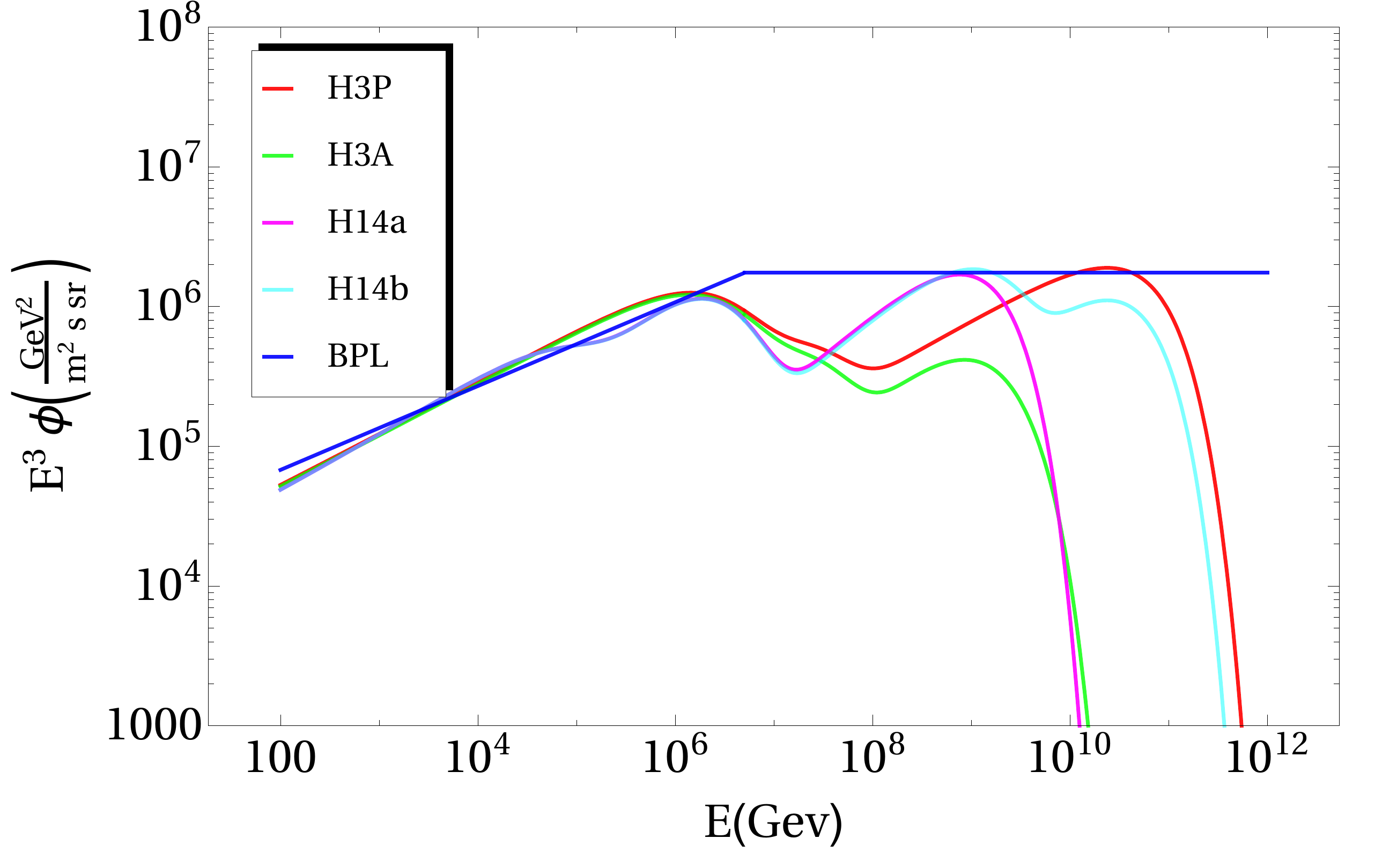}
\caption{\small Comparison between the parameterisations of the cosmic
  ray flux used in this work.}
\label{fig:CRcomparison}
\end{figure}

We consider two types of `all-proton' spectra, one where the third
extra-galactic population contains contributions from all 5 nuclear
groups, and another where only protons contribute, denoted
respectively by {H3A} and {H3P}~\cite{Gaisser:2011cc}.  These
parameterisations have been
extended~\cite{Stanev:2014mla,Gaisser:2013bla} to include both
additional heavy nuclear species, H14a, and to include a fourth
population, H14b.

All the above spectra are compared in figure~\ref{fig:CRcomparison} with
the flux rescaled by $E^3$ so that the region above the `knee' is a
horizontal line for the BPL spectrum and the difference from the other
more recent parameterisations is emphasised.  The latter are similar
up to $\sim 10^{8}$ GeV, but differ significantly thereafter.
Although now outdated, the results with the BPL spectrum are required
for comparison with older calculations of the prompt neutrino flux.

\subsection{The cascade equations and their solution}

We now review the cascade formalism in the framework of the $Z$-moment
approach \cite{Lipari:1993hd,Gaisser:1990vg} which is used to simulate
the propagation of high energy particles and their decay products
through the atmosphere.  The aim is to solve a series of coupled
differential equations dependent on the slant depth $X(l, \theta)$
measuring the atmosphere traversed by a particle:
\begin{equation*}\label{eq:X}
X(l, \theta) \equiv \int_l^\infty \rho[h(l', \theta)] \mathrm{d}l' \, ,
\end{equation*}
where $\rho$ is the density of the atmosphere dependent on the
distance from the ground $l$ (along the particle trajectory) as well
as on the zenith angle $\theta$.  We adopt an isothermal model of the
atmosphere, appropriate for atmospheric depths 10--40 km within which
the bulk of particle production occurs:
\begin{equation}
\rho(h) = \rho_{0} \exp \left( -h/h_0 \right) \, , \quad \rho_0 = 2.03 \times 10^{-3}\,\,{\rm gm~cm}^{-3} \, , 
\quad h_{0} = 6.4 \,\,{\rm km} \, .  
\end{equation} 
The horizontal depth of the atmosphere is
$X \simeq 3.6 \times 10^{4}~{\rm gm~cm}^{-2}$ while its vertical depth
is $\simeq 1.3 \times 10^{3}~{\rm gm~cm}^{-2}$.  As in previous
calculations, we are concerned with small angles about the vertical,
$\theta \simeq 0$, where the conventional atmospheric neutrino flux
arising from the decays of charged pions and kaons is the smallest.

For a particle of species $j$ and energy $E_j$ that has traversed a
slant depth $X$, the cascade equation for the corresponding flux
$\phi_{j}(E_j,X)$ is
\begin{equation}
  \label{eq:cascadeeq}
  \frac{\mathrm{d}\phi_j}{\mathrm{d}X} = - \frac{\phi_j}{\lambda_j} - \frac{\phi_j}{\lambda_j^\text{dec}} +
  \sum_k S_{kj}(E_j, X) \, ,
\end{equation}
where $\lambda_{j}$ is the interaction length, $\lambda_{j}^{\rm dec}$
is the decay length, and $S_{kj}$ are `(re)generation functions'
describing the production of particle $j$ from particle $k$ (where the
sum includes $k=j$).  This says that as a particle traverses the
atmosphere, its flux will decrease when the particle undergoes an
interaction (thus losing energy) or decays, as well as increase from
the decay or interaction of other particle species.  The
(re)generation function is:
\begin{equation}
S_{kj}(E_j,X) = \int_{E_j}^\infty \frac{\phi_{k}(E_{k}^{'},X)}{\lambda_{k}(E_{k}^{'})} \frac{\mathrm{d}n(k \rightarrow j; E^{'}_k, E_j)}{\mathrm{d}E_j} \mathrm{d}E_k^{'} \, ,
\end{equation}
where $\mathrm{d}n(k \rightarrow j; E^{'}_k, E_j)$ is the differential
transition rate between particle species $k$ and $j$.  Assuming that
the particle flux factorises into components dependent respectively on
the energy $E$ and the slant depth $X$,
\begin{equation} 
\phi_{k}(E,X)\equiv \phi_{k}(E) \times \widetilde{\phi}_{k}(X) \, , 
\end{equation} 
it can be rewritten more simply as 
\begin{equation}
S_{kj}(E_j,X) =
\frac{\phi_{k}(E_j,X)}{\lambda_{k}(E_j)}Z_{kj}(E_j) \, ,
\end{equation}
with the key property that the moment $Z_{kj}$,
\begin{equation}
\label{eq:Zprod}
Z_{k j}(E_j) = \int_{E_j}^\infty \frac{\phi_{k}(E_{k}^{'},X)}{\phi_{k}(E_{j},X)} \frac{\lambda_{k}(E_{j})}{\lambda_{k}(E_{k}^{'})} \frac{\mathrm{d}n(k \rightarrow j; E^{'}_k, E_j)}{\mathrm{d}E_j} dE_k^{'} \, ,
\end{equation}
is independent of the slant depth $X$ (which cancels in the ratio of
fluxes).

Under this factorisation assumption, the cascade equations describing
the flux of the various relevant species (protons $p$, mesons $m$, and
leptons $l$) as they propagate through the atmosphere can be written
as a set of coupled differential equations:\footnote{Here `meson'
  includes charmed baryons such as $\Lambda_c^{\pm}$ which also yield
  a prompt neutrino flux.}
\begin{align}
\label{eq:Ncasc} 
\frac{\mathrm{d}\phi_{p}}{\mathrm{d}X} &= -\frac{\phi_{p}}{\lambda_{p}} + Z_{pp}\frac{\phi_{p}}{\lambda_{p}}\, ,\\
\label{eq:Mcasc} 
\frac{\mathrm{d}\phi_{m}}{\mathrm{d}X} &= - \frac{\phi_{m}}{\rho \mathrm{d}_{m}(E)} - \frac{\phi_{m}}{\lambda_{m}} + Z_{mm}\frac{\phi_{m}}{\lambda_{m}} + Z_{pm}\frac{\phi_{p}}{\lambda_{p}} \, ,\\
\label{eq:lcasc} 
\frac{\mathrm{d}\phi_{l}}{\mathrm{d}X} &= \sum_{m} Z_{m \rightarrow l}\frac{\phi_{m}}{\rho d_m} \, ,
\end{align}
where in the last equation the sum is restricted to the charmed
hadrons that contribute to the prompt flux. Here
$d_{m}(E) = c \beta\gamma\tau_{m}$ is the decay length of a particle
with proper lifetime $\tau_{m}$.

Although the solution of these equations is in general quite involved,
there are simple asymptotic solutions.  The first equation for the
proton flux~(\ref{eq:Ncasc}) can be trivially integrated to give
\begin{equation} 
 \label{eq:phiN}
  \phi_{p}(E,X) = \phi^{(0)}_p (E) \exp \left(-X / \Lambda_p (E) \right) \, ,
\end{equation}
where we have defined the nucleon attenuation length as
\begin{equation}
\Lambda_{p}(E) \equiv \lambda_{p}(E)/(1-Z_{pp}(E)) \, .
\end{equation}
This depends in general on the nucleon's interaction length in the
atmosphere $\lambda_{p}(E)$:
\begin{equation}
\lambda_{p}(E)= \langle A \rangle/N_{0} \sigma_{pA}(E) \, ,
\end{equation}
where $\langle A \rangle = 14.5$ is the average atomic number of air
molecules, $N_0$ is Avogadro's number, and the total inelastic
air-nucleon cross-section is denoted by $\sigma_{pA}$.

Concerning the total proton-air cross-section, several
parameterisations are
available~\cite{Kalmykov:1993qe,Mielke:1994un,Bugaev:1998bi,Heck:1998vt,Ahn:2009wx}.
We use the {\tt QGSJet0.1c} model~\cite{Mielke:1994un} which fits
available data well through the relevant energy range, including
recent measurements made at the LHC~\cite{Antchev:2011vs} and the
Pierre Auger Observatory~\cite{Collaboration:2012wt}.  The prompt
neutrino flux in fact depends very weakly on the modelling of
$\sigma_{pA}(E)$~\cite{Garzelli:2015psa}.

Given eq.~(\ref{eq:phiN}) the cascade equation~(\ref{eq:Mcasc}) for
the meson flux can be solved by neglecting, in the low energy limit,
the interaction and regeneration terms and, in the high energy limit,
the decay terms.  This yields two asymptotic solutions:
\begin{equation}
  \label{eq:mesonlow}
  \phi_{m}^{\rm low}(E) = \phi_{p}^{(0)}(E) \frac{Z_{pm}(E)}{\Lambda_{p}(1 - Z_{pp})} \rho d_{m}
  e^{-X/\Lambda_p}\, ,
\end{equation}
\begin{equation}
  \label{eq:mesonhigh}
\phi_{m}^{\rm high}(E)=  \phi_{p}^{(0)}(E) \frac{Z_{pm}(E)}{(1-Z_{pp})}\frac{(e^{-X/\Lambda_m} -e^{-X/\Lambda_p})}{1 - \Lambda_{p}/\Lambda_{m}}\, ,
\end{equation}
where the dependence on the energy of the attenuation lengths
$\Lambda_p$ and $\Lambda_m$ is implicit.  Because of the additional
dependence on the decay length in the low energy solution, these
fluxes effectively scale with the proton flux~(\ref{eq:phiN}) as
follows:
\begin{align}
  \label{eq:scale1}
&\phi_{m}^{\rm low}(E) \propto E\phi_{p}(E)\, ,\\
\label{eq:scale2}
&\phi_{m}^{\rm high}(E) \propto \phi_{p}(E)\, .
\end{align}
These intermediate solutions for meson fluxes are subsequently used as
inputs in the corresponding low and high energy solutions for the
leptonic decay of a meson $m \rightarrow l$, with either $l = \mu$ or
$\nu$.  The final vertical flux of leptons expected at the detector
can then also be described by two asymptotic solutions, valid in the
low-- and high--energy regimes respectively:
\begin{align}
\label{eq:phil1}
\phi^{\rm low}_{l,m}(E) &= \phi_{p}^{(0)}(E) \frac{Z_{pm}(E)}{1-Z_{pp}} Z^{\rm low}_{ml}(E)\, , \\
\label{eq:phil2}
\phi^{\rm high}_{l,m}(E) &= \phi_{p}^{(0)}(E) \frac{\epsilon_{m}}{E} \frac{Z_{pm}(E)}{1-Z_{pp}}\frac{\ln(\Lambda_{m}/\Lambda_{p})}{1-\Lambda_{p}/\Lambda_{m}} Z^{\rm high}_{ml}(E) \, . 
\end{align}
where $\epsilon_{m}$ is a critical energy below which the probability
of a meson to decay is greater than it is to interact:
\begin{equation}
  \label{eq:critical}
\epsilon_{m} = \frac{m_{m}c^{2}h_{0}}{c\tau_{m} \cos\theta} \, .
\end{equation}
The smaller the critical energy, the longer the decay length, hence the
more energy the particle will lose by interactions in the atmosphere
before it actually decays. In eqs.~(\ref{eq:phil1})
and~(\ref{eq:phil2}), $\phi_{l,m}$ represents the flux of
lepton $l$ from the decays of the meson $m$.

Pions and kaons have a critical energy of $\mathcal{O}(10^2)$ GeV.
However, heavy quark mesons, such as $B$ and $D$, are characterised by
much larger critical energies of $\mathcal{O}(10^{7})$ GeV and
therefore mainly decay before losing energy in interactions with the
atmosphere. This is why the `prompt' neutrino flux from their decays
is expected to dominate over the `conventional' flux from $\pi, K$ at
high energies. The contribution of $B$ mesons is usually neglected
(despite a larger critical energy as compared to $D$ mesons) because
the $b$-quark pair production cross-section is $<10\%$ of that of
$c$-quark pairs up to $\sim 100$ PeV.  However at such high energies,
the contribution from D mesons would be damped as they start
interacting before decaying, hence we show the prompt neutrino flux
only up to $10^{7.5}$ GeV,

The final step in solving the cascade equations in the $Z$-moment
approach is the geometrical interpolation of the low-- and
high--energy asymptotic solutions~(\ref{eq:phil1})
and~(\ref{eq:phil2}) which yields the final expression for the prompt
lepton (neutrino) flux:
\begin{equation}
  \label{eq:int}
  \phi_{l}(E)=\sum_{m}\frac{\phi_{l,m}^{\rm low}(E) \times
    \phi_{l,m}^{\rm high}(E)}{\phi_{l,m}^{\rm low}(E) + \phi_{l,m}^{\rm high}(E)} \, .
\end{equation}
In the sum over mesons contributing to the prompt flux we include (the
leptonic decays of) $D^{0}$, $\bar{D}^{0}$, $D^{\pm}$, $D_{s}^{\pm}$
and $\Lambda_{c}^{\pm}$.  In fact $D^{0}$, $\bar{D}^{0}$, and
$D^{\pm}$ account for the bulk of charm production in the atmosphere,
with the other mesons contributing only around 10\%.

\subsection{Computation of the $Z$ moments}

We need to compute the various $Z$-moments in order to evaluate the
prompt flux~(\ref{eq:int}), of which the most crucial is the nucleon
to meson moment, $Z_{pm}$, which depends on the charm production
cross-section in $pp$ collisions.  We now discuss how to compute
$Z_{ml}$, $Z_{pp}$, $Z_{mm}$ and $Z_{pm}$, and the corresponding
theoretical uncertainties.

The generic $Z$-moment~(\ref{eq:Zprod}) defined earlier can be written
for a (re)generation moment accounting for the interaction of a proton
or a meson with an air nucleus, as
\begin{equation}
\label{eq:Zprod3}
Z_{k j}(E_j) = \int_{E_j}^\infty \frac{\phi_{k}(E_{k}^{'},X)}{\phi_{k}(E_{j},X)} \frac{\lambda_{k}(E_{j})}{\lambda_{k}(E_{k}^{'})} \frac{\mathrm{d}n(kA \rightarrow j; E^{'}_k, E_j)}{\mathrm{d}E_j} \mathrm{d}E_k^{'} \, ,
\end{equation}
while the decay moments that account for the leptonic decays of mesons
are given by
\begin{equation}
\label{eq:ZM}
Z_{m l}(E_l) = \int_{E_l}^\infty \frac{\phi_{m}(E_{k}^{'},X)}{\phi_{m}(E_{l},X)}
\frac{d_{m}(E_l)}{d_{m}(E_k^{\prime})}
\frac{\mathrm{d}n(k \rightarrow l+X;E_k^{\prime},E_l)}{\mathrm{d}E_l} \mathrm{d}E_k^{'}  \, .
\end{equation}
Here the differential distributions
$\mathrm{d}n(i \rightarrow f; E^{\prime}, E)/\mathrm{d}E$ correspond
to the number $n$ of final state particles $f$ with energies between
$E$ and $E + \mathrm{d}E$ produced in an interaction where the initial
state particle has energy $E^{\prime}$.

We now outline how each of the moments has been computed in this work:

\bigskip \noindent $\bullet$ For the calculation of the leptonic decay
moment $Z_{ml}$~(\ref{eq:ZM}), we use the fact that the energy
distribution of leptons from charmed meson decays obeys a scaling law:
\begin{equation}
  \mathrm{d}n(m \rightarrow l+X;E^{\prime},E) =
  F_{m \rightarrow l}\left(\frac{E}{E^{\prime}}\right)\frac{\mathrm{d}E}{E^{\prime}} \, ,
\end{equation}
where $F_{m\to l}(E)$ is the energy spectrum of the lepton $l$ from
the decay of the meson $m$, computed in the rest frame of the latter.
Defining the scaling variable $x_{E} = E/E^{\prime}$, we obtain
\begin{equation}
\label{eq:ZM2}
Z_{ml}(E) = \int_{0}^{1} \mathrm{d}x_{E} \frac{\phi_{m}(E/x_{E})}{\phi_{m}(E)}F_{m \rightarrow l}(x_{E}) \, .
\end{equation}
Exploiting the fact that the meson flux $\phi_m(E_m)$ scales in the
high--energy~(\ref{eq:scale1}) and low--energy~(\ref{eq:scale2})
limits we find that for the BPL cosmic ray spectrum, the leptonic
moment reduces to a relatively simple expression
\begin{equation}
  \label{eq:leptonZ}
Z_{ml}(E_l) = \int_{0}^{1} \mathrm{d}x_{E}\,\, x_{E}^{\beta}\,\,F_{m \rightarrow l}(x_{E}) \, ,
\end{equation}
where the exponent is $\beta^{\rm low} = \lbrace1.7,\,2 \rbrace$ in
the low--energy solution, and
$\beta^{\rm high}=\lbrace 2.7,\,3 \rbrace$ in the high-energy
solution, for energies below and above the `knee' respectively.

The calculation of the leptonic energy spectrum $F(x_{E})$ from
charmed meson decays is performed with the {\tt
  Pythia8}~\cite{Sjostrand:2014zea} event generator and a boost is
applied to transform $F(x_{E})$ from the laboratory to the meson rest
frame. For the leptonic branching fractions of charmed mesons, we use
the Particle Data Group recommended values ~\cite{Agashe:2014kda} for
inclusive decays:
$\mathcal{B} (D^{\pm} \rightarrow \nu_{l} X) = 0.161$,
$\mathcal{B} (D^{0} \rightarrow \nu_{l} X) = 0.065$,
$\mathcal{B} (D^{\pm}_{s} \rightarrow \nu_{l} X) = 0.065$, and
$\mathcal{B} (\Lambda_{c}^{\pm} \rightarrow \nu_{l} X) = 0.028$.
These values are adopted for both muon and electron neutrinos. The
uncertainty on the branching fractions is well below 10\% for $D^0$
and $D^{\pm}$, which are the most abundantly produced hadrons due to
their large fragmentation fractions. Our result for the $Z_{ml}$
moments using the BPL cosmic ray spectrum are quite consistent with
those reported earlier~\cite{Gondolo:1995fq}.

\begin{figure}[t]
\centering 
\includegraphics[scale=0.28]{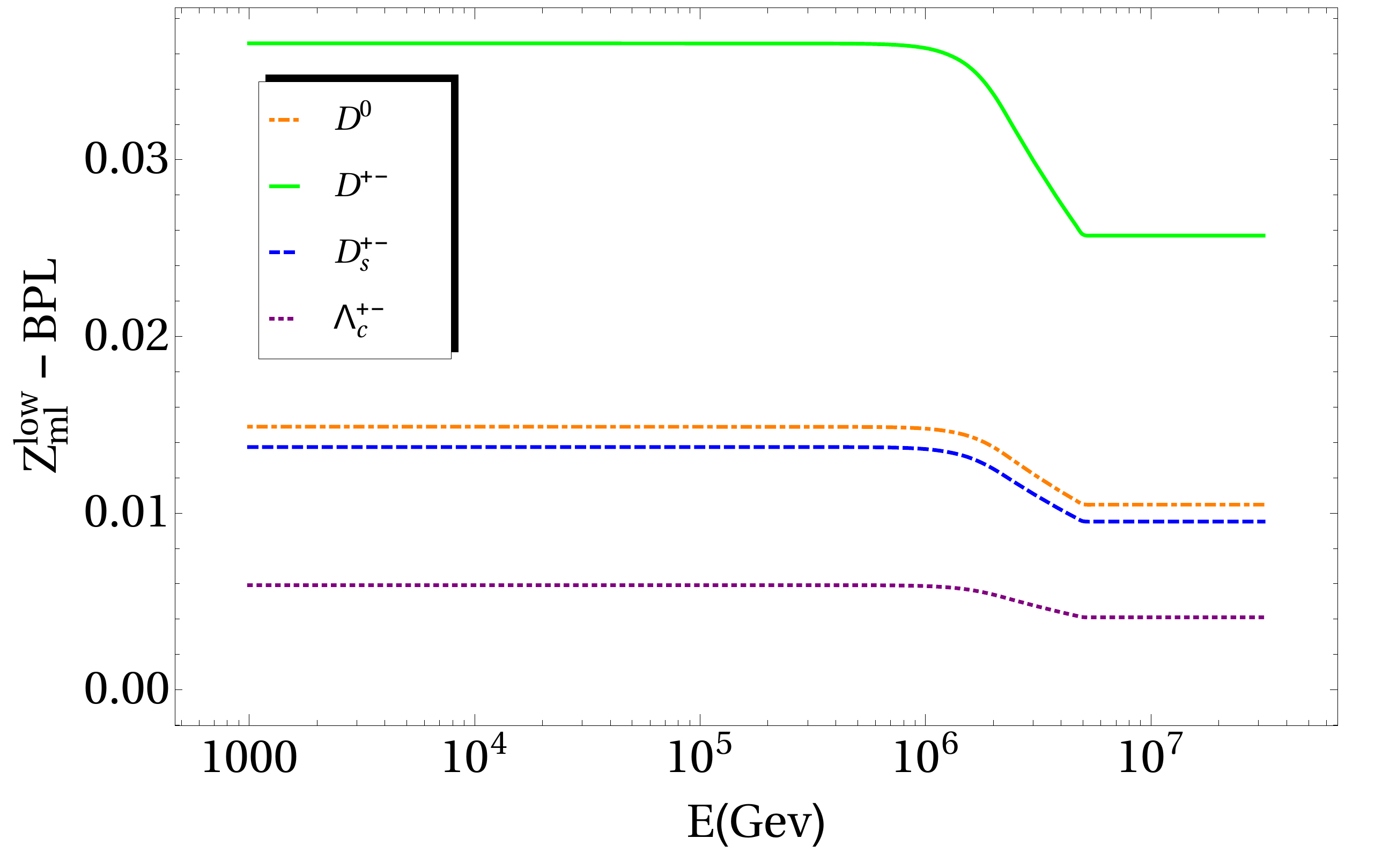}
\includegraphics[scale=0.28]{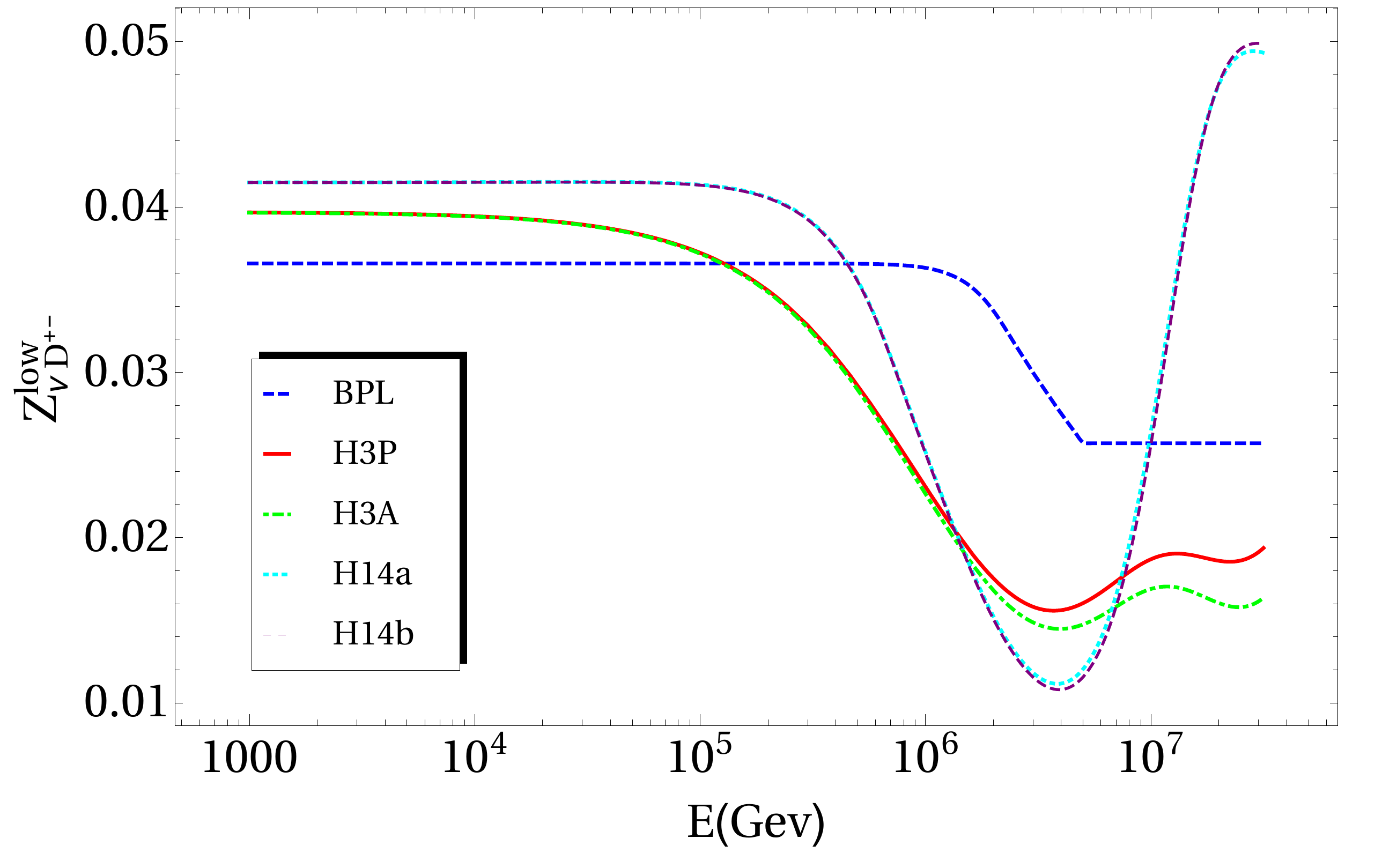}
\caption{\small Left: Comparison of the leptonic moment
  $Z^{\rm low}_{ml}(E)$, assuming the BPL cosmic ray flux, for the
  four charmed mesons that contribute mainly to the prompt flux.
  Right:
  Comparison of the moment $Z^{\rm low}_{ml}(E)$ for $D^{\pm}$ mesons
  only for the 5 different parameterisations of the cosmic ray flux.}
\label{fig:Zhl_comparison}
\end{figure}

In figure~\ref{fig:Zhl_comparison} we compare the low-energy solution
for the leptonic moment $Z^{\rm low}_{ml}(E)$ using the BPL cosmic ray
spectrum for the four charmed mesons that contribute to the prompt
flux, and where a sum over charge conjugate states is understood.
Note that the decays of $D^0$ and $D^{\pm}$ contribute the bulk of the
prompt leptonic flux. We also show a comparison of
$Z^{\rm low}_{ml}(E)$ for $D^{\pm}$ only, using the different
parameterisations of the cosmic ray flux to illustrate the large
variations.
 
\bigskip \noindent 
$\bullet$ When calculating the regeneration moments $Z_{pp}$ and
$Z_{mm}$ that account for the interactions of protons and mesons in
the atmosphere yielding a final state containing the same particle
species, we follow previous studies
\cite{Bhattacharya:2015jpa,Garzelli:2015psa} in adopting scaling laws
for the proton-proton and meson-proton cross-sections, viz.
\begin{align}
\label{eq:Zpp}
\frac{\mathrm{d}\sigma(pA \to p+X,E^{\prime},E)}{\mathrm{d}x_{E}} &\simeq \sigma_{pA}(E) (1 + n_{1}) (1-x_{E})^{n_{1}} \, ,\\
\label{eq:Zhh}
\frac{\mathrm{d}\sigma(mA \to m+X,E^{\prime},E)}{\mathrm{d}x_{E}} &\simeq A^{3/4} \sigma_{Kp}(E) (1 + n_{2}) (1-x_{E})^{n_{2}} \, ,
\end{align}
where, as before, $x_{E} = E/E^{\prime}$ is the fraction of the
original energy retained by the incoming particle after interaction
with an air nucleus in the atmosphere and the exponents are
$n_{1} = 0.51$ and $n_{2} = 1.0$

Eq.~(\ref{eq:Zhh}) is based on the approximation that the
cross-section for charmed meson scattering off nucleons can be related
to the corresponding kaon-proton cross-section. The attenuation length
of charmed mesons will be given under the same approximation as
\cite{Pasquali:1998ji}
\begin{equation}
  \label{eq:lambdaH}
\Lambda_{m}(E) \simeq \frac{A}{N_{0}\sigma_{pA}(E)} \frac{\sigma_{pp}(E)}{\sigma_{Kp}(E)} \frac{1}{(1-Z_{KK})} \, ,
\end{equation}
where the dependence of the kaon-proton cross-section on energy is
from \cite{Agashe:2014kda}.

\begin{figure}[t]
\centering 
\includegraphics[scale=0.28]{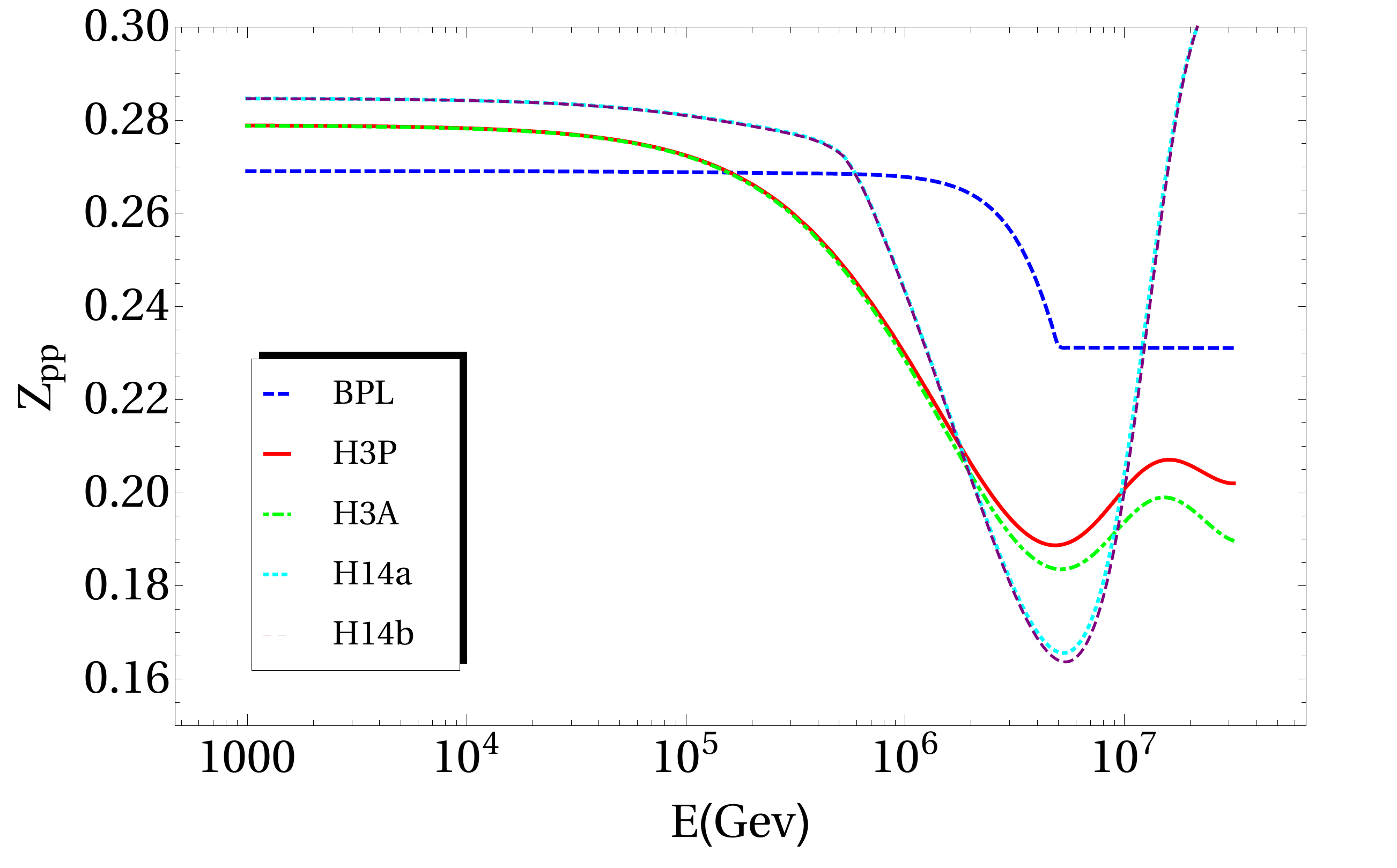}
\includegraphics[scale=0.28]{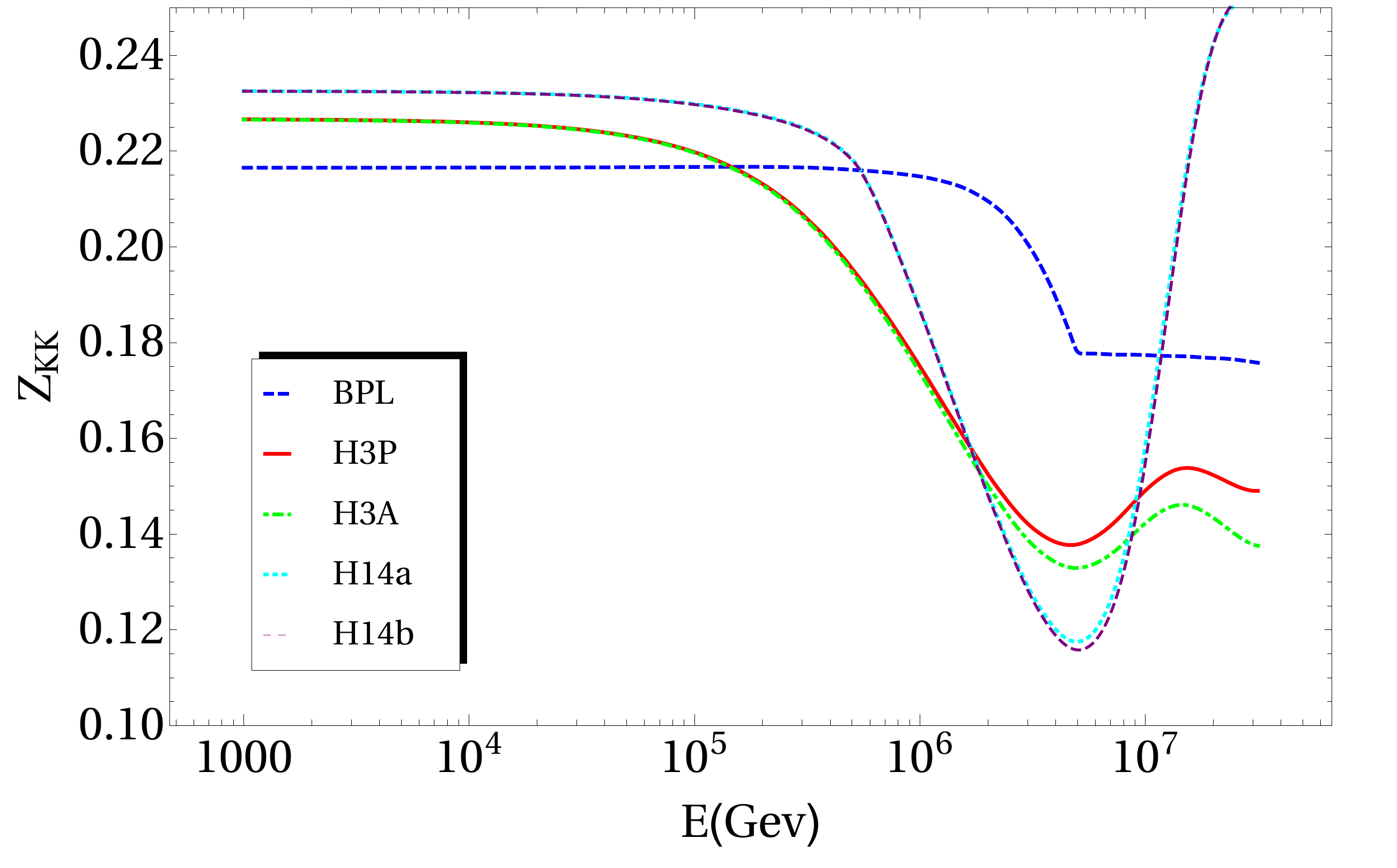}
\caption{\small Left: Comparison of the proton regeneration moment
  $Z_{pp}(E)$ for the 5 cosmic ray flux parameterisations.  Right:
  Same comparison, now for the meson regeneration moments
  $Z_{KK}(E)$.}
\label{fig:Zpp_comparison}
\end{figure}

In figure~\ref{fig:Zpp_comparison} we compare the proton and meson
regeneration moments $Z_{pp}(E)$ and $Z_{KK}(E)$ for the 5 cosmic ray
flux parameterisations.  As for the leptonic moments, differences
become appreciable only at high energies above the `knee'.
 
\bigskip \noindent $\bullet$ Finally we discuss the calculation of the
proton-meson moment $Z_{pm}$, which is the main ingredient of the
present work, as it contains the information on charm production in
high energy collisions.  The number distribution can be related to the
differential charm production cross-section as:
\begin{equation}
\frac{\mathrm{d}n(pA\rightarrow h+X;E^{\prime},E)}{\mathrm{d}E} = \frac{1}{\sigma_{pA}(E^{\prime})}\frac{\mathrm{d}\sigma(pA\rightarrow h+X;E^{\prime},E)}{\mathrm{d}E} \, .
\end{equation}
We assume that the charm production cross-section scales with the mean
atomic number of air $\langle A\rangle$ as compared to the
corresponding $pp$ cross-section
\begin{equation}
\sigma(pA \rightarrow D+X) \simeq \langle A\rangle \, \sigma(pp \rightarrow D+X) \,, 
\end{equation}
where $D$ is a generic charmed meson.  This approximation is justified
since even for forward $D$ production in $pPb$ collisions, the nuclear
modification of the differential $D$ hadron cross-section results in a
suppression of at most 10\%~\cite{Gauld:2015lxa}.  Although such
effects are expected to increase in strength with atomic number, it is
reasonable to ignore them when air is the target.  This approximation
is also supported by recent $B$ production data on $pPb$ collisions at
the LHC~\cite{Khachatryan:2015uja} which show no evidence for nuclear
modification effects.

Since we assume that ratios of fluxes are independent of the slant
depth $X$ to first approximation, we can write down a simplified
version of the $Z_{pm}$ moment~(\ref{eq:Zprod3}) in terms of the charm
production cross-section as follows:
\begin{equation}
\label{eq:Zph}
Z_{pm}(E_m) = \int_{E_m}^\infty \frac{\phi_{p}(E_{p}^{'})}{\phi_{p}(E_{m})}
\frac{\langle A \rangle }{\sigma_{pA}(E_{m})}
\frac{\mathrm{d}\sigma(pp \rightarrow D+X; E^{'}_p, E_m)}{\mathrm{d}E_m} \mathrm{d}E_p^{'} \, ,
\end{equation}
Our calculation of the differential charm production cross-section in
$pp$ collisions at high energies has been discussed in detail
earlier~\cite{Gauld:2015yia}.  As explained therein, it is based on
perturbative QCD as implemented in the NLO Monte Carlo event generator
{\tt POWHEG}~\cite{Alioli:2010xd}, benchmarked with the corresponding
{\tt FONLL}~\cite{Cacciari:2001td} and {\tt
  aMC@NLO}~\cite{Alwall:2014hca} calculations.  The input PDF set is
NNPDF3.0+LHCb~\cite{Ball:2014uwa,Gauld:2015yia}, which includes the
constraints on the small-$x$ gluon from the LHCb 7 TeV charm
production cross-sections.  The parton showering and fragmentation are
modeled with {\tt Pythia8}~\cite{Sjostrand:2014zea} using the Monash
2013 tune~\cite{Skands:2014pea}.  This is consistent with the
semi-analytical fragmentation implemented in {\tt FONLL}, tuned to LEP
data~\cite{Cacciari:2005uk}. For the fragmentation probabilities,
which describe the transition $f(c\to D)$ for the different types of
charmed mesons, rather than using the default {\tt Pythia8} values we
use the recent LHCb measurements~\cite{Aaij:2013mga}:
$f(c\to D^0)=0.565$, $f(c\to D^{\pm})=0.246$,
$f(c\to D_s^{\pm})=0.080$, and $f(c\to \Lambda_c)=0.094$.

\begin{figure}[t]
\centering 
\includegraphics[scale=0.28]{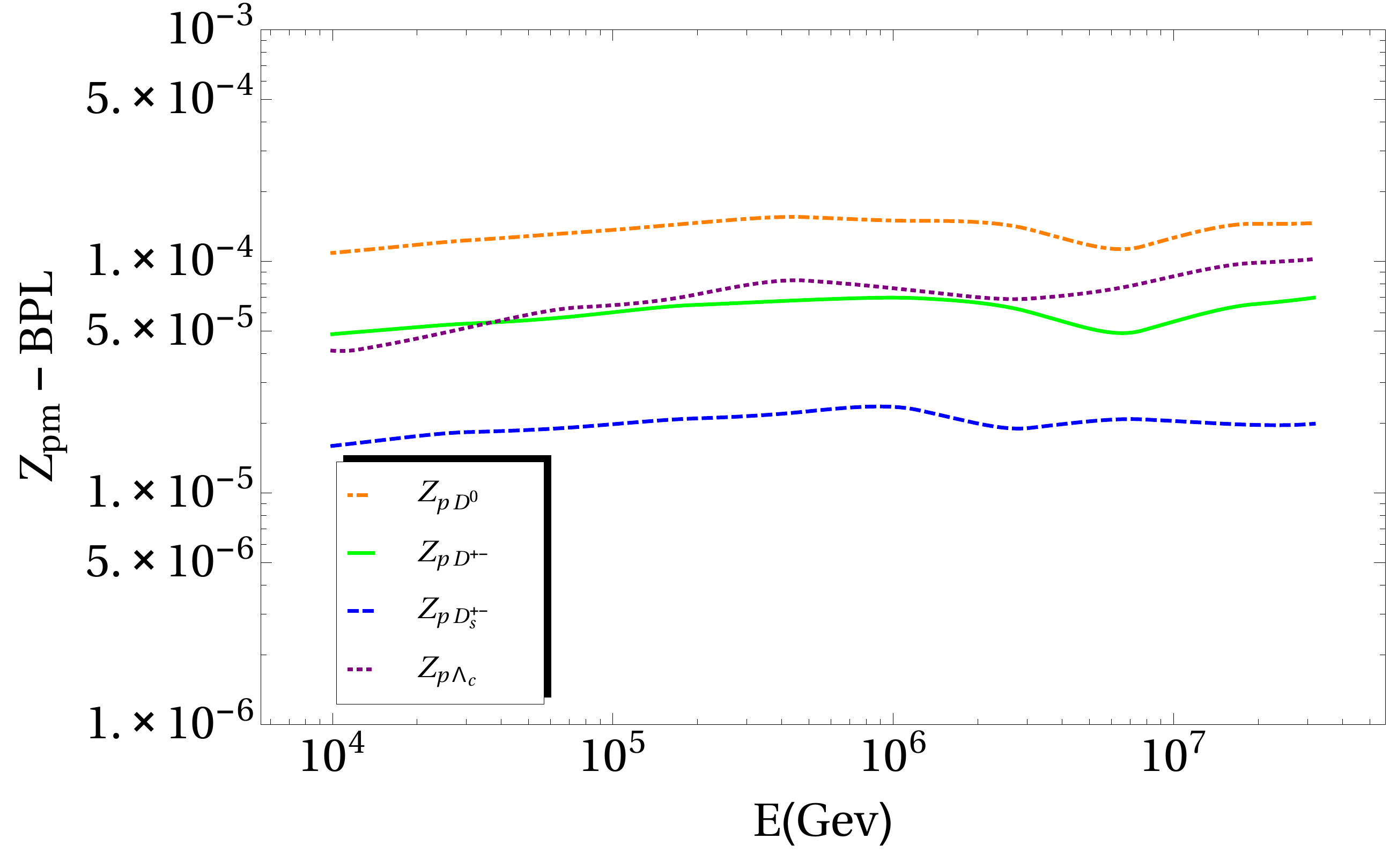}
\includegraphics[scale=0.28]{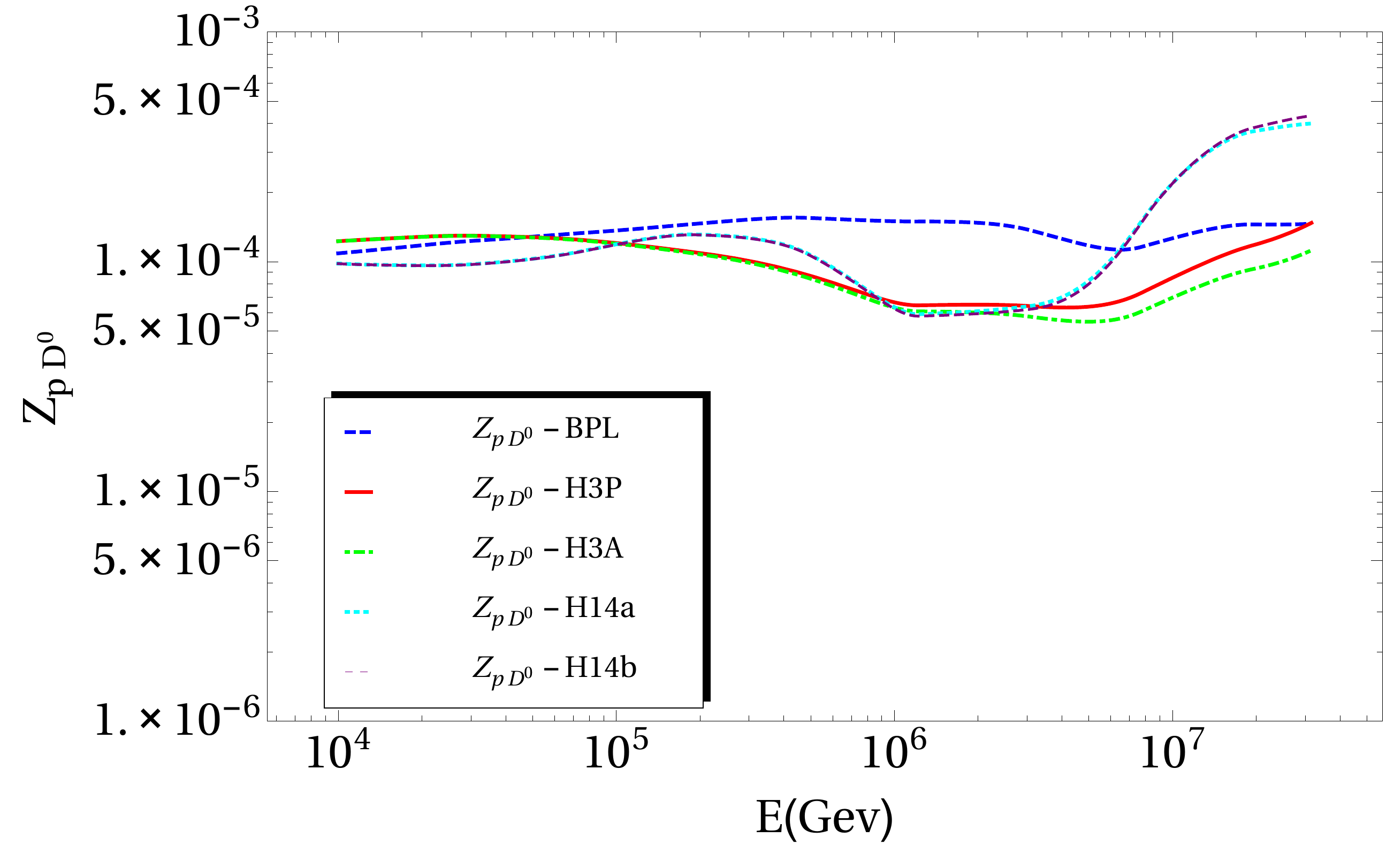}
\caption{\small Same as figure~\ref{fig:Zhl_comparison} for the
  proton-meson moment $Z_{pm}(E)$.}
\label{fig:Zph_comparison}
\end{figure}

Using this framework, we have computed the moment $Z_{pm}(E)$ for a
wide range of energies, from $10^3$ to $10^{7.5}$ GeV.  This requires
the calculation of the charm production cross-section for incoming
proton energies up to $E_p=10^{10.5}$ GeV in the laboratory
frame.\footnote{ The upper integration limit in $Z$-moments such as
  eq.~(\ref{eq:Zph}) is actually a fixed value $E^{\rm max}_p$ rather
  than infinity.  We have verified that provided that this upper
  integration limit is at least about 100 times larger than $E_{m}$,
  the numerical results are insensitive to the specific choice for
  $E^{\rm max}_p$.}  The {\tt POWHEG} calculation is done in the
center-of-mass frame for a wide range of $\sqrt{s}$ values, then
boosted to the laboratory frame. In each case we have computed all the
associated theoretical uncertainties from missing higher-orders, PDFs,
and from the value of the charm mass~\cite{Gauld:2015yia} as follows:

\begin{itemize}

\item The charm quark pole mass is varied as $m_c = (1.5\pm0.2)$~GeV,
	
\item Renormalisation and factorisation scales are varied
  independently by a factor of 2 around the central scale
  $\mu_0 = \sqrt{p_T + m_c^2}$, with the constraint
  $1/2 < \mu_F/\mu_R < 2$.
	
\item PDF uncertainties are included at 68\% CL,
	
\item Finally the total theory uncertainty is obtained by addition in
  quadrature of these 3 components so may be considered as a crude
  `$1\sigma$' band.
	
\end{itemize}
As with the other moments, the calculation of $Z_{pm}(E)$ is performed
for all 5 cosmic ray flux parameterisations.
In figure~\ref{fig:Zph_comparison} we show the central theory
prediction for $Z_{pm}(E)$ for the BPL spectrum for the four relevant
charmed mesons (left plot) and then, for the $D^0$ and $\bar{D}^{0}$
mesons only, using all parameterisations (right plot).

\section{Results}
\label{sec:results}

This section contains our main result, the updated calculation of the
prompt neutrino flux.  We discuss its dependence on the various
inputs, in particular the adopted cosmic ray flux parameterisation and
PDF set used.  We compare our result with other recent calculations
and also provide the spectral index of the prompt flux as a function
of energy.

\subsection{The prompt neutrino flux}

\begin{figure}[t]
\centering 
\includegraphics[scale=0.4]{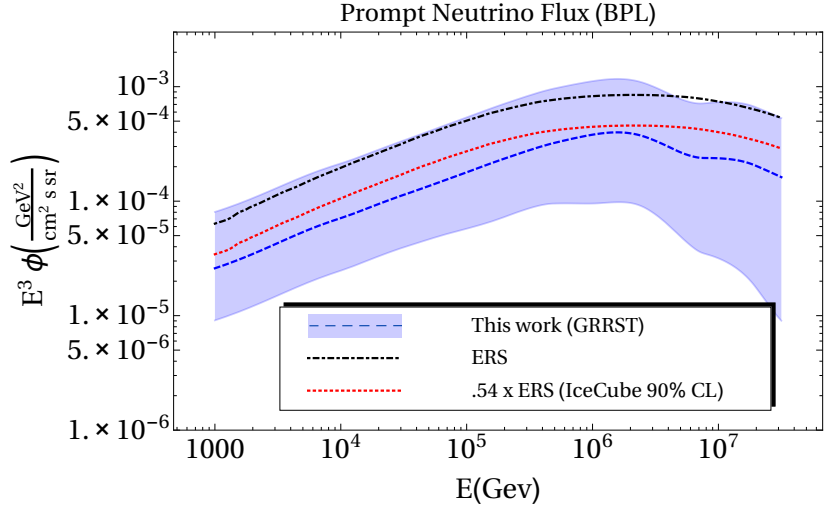}
\vspace{-1cm}
\caption{\small The prompt neutrino flux using the BPL cosmic ray
  spectrum as input.  The error band includes all relevant sources of
  theoretical uncertainties: from PDFs (68\% CL), missing higher
  orders and the charm mass, as discussed in the text.  The ERS
  benchmark calculation \cite{Enberg:2008te} is shown for comparison,
  as is the recent 90\% CL upper limit on the prompt flux from IceCube
  \cite{Schoenen:2015ipa}.}
\label{fig:flux1}
\end{figure}

Figure~\ref{fig:flux1} shows the prompt neutrino flux up to $10^{7.5}$
GeV using the BPL cosmic ray spectrum. Since PDF uncertainties have
been substantially reduced using the LHCb data, the error band is
dominated by the `scale uncertainties' of the NLO perturbative QCD
calculation which can be reduced only when the corresponding NNLO
result is available~\cite{Czakon:2015owf}.  However at energies above
a PeV, PDF uncertainties still make an important contribution to the
total error band.  We also show the central value of the ERS
calculation~\cite{Enberg:2008te}, which has been used as a benchmark
in several IceCube analyses but is now in tension with the 90\% CL
upper limit labeled `0.54$\times$ERS' \cite{Schoenen:2015ipa}. The
central value of our calculation is a factor of 2 smaller, and just
below the IceCube limit on the prompt neutrino flux. Note that this
limit should be interpreted with some care, since it depends e.g. on
the specific parameterisation of the cosmic ray flux in the analysis.

\begin{figure}[t]
\centering 
\includegraphics[scale=0.40]{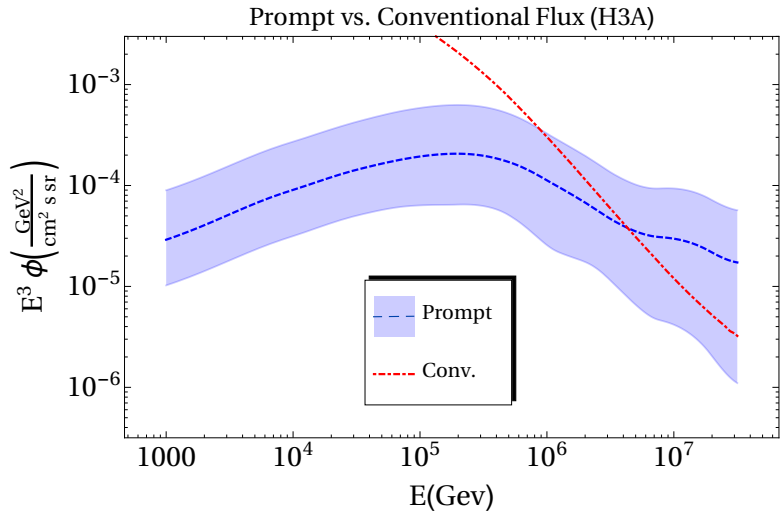}
\vspace{-1cm}
\caption{\small The prompt neutrino flux and its uncertainty using the
  H3A cosmic ray spectrum as input, compared to the conventional
  neutrino flux at IceCube \cite{Aartsen:2015xup}.}
\label{fig:conventional}
\end{figure}

In figure~\ref{fig:conventional} we compare the prompt neutrino flux
with the conventional neutrino flux from the decays of pions and
kaons, using the same cosmic ray spectrum (H3A).  We use the updated
calculation~\cite{Honda:2011nf} for the South Pole location as
implemented in the code {\tt NeutrinoFlux} used by the IceCube
collaboration \cite{Aartsen:2015xup} . Whereas for the conventional
flux the location of the experiment is important (as this determines
the geomagnetic rigidity cut-off which filters incoming cosmic rays),
this is irrelevant for the prompt flux which arises from the
interaction of much higher energy cosmic rays). The cross-over energy
where the prompt component begins to exceed the conventional one is
about $4\times 10^6$ GeV.
  
As discussed in \S.~\ref{sec:crflux}, an essential component of any
calculation of the prompt neutrino flux is the parameterisation of the
incoming cosmic ray flux, which is rather uncertain at the relevant
high energies. Since cosmic rays with energies
$\gtrsim (100-1000)E_{\nu}$ contribute to the prompt neutrino flux at
a given $E_{\nu}$, a prediction of the prompt flux up to $10^{7.5}$
GeV requires knowledge of the cosmic ray flux up to at least
$10^{10.5}$ GeV.

\begin{figure}[t]
\centering 
\includegraphics[scale=0.27]{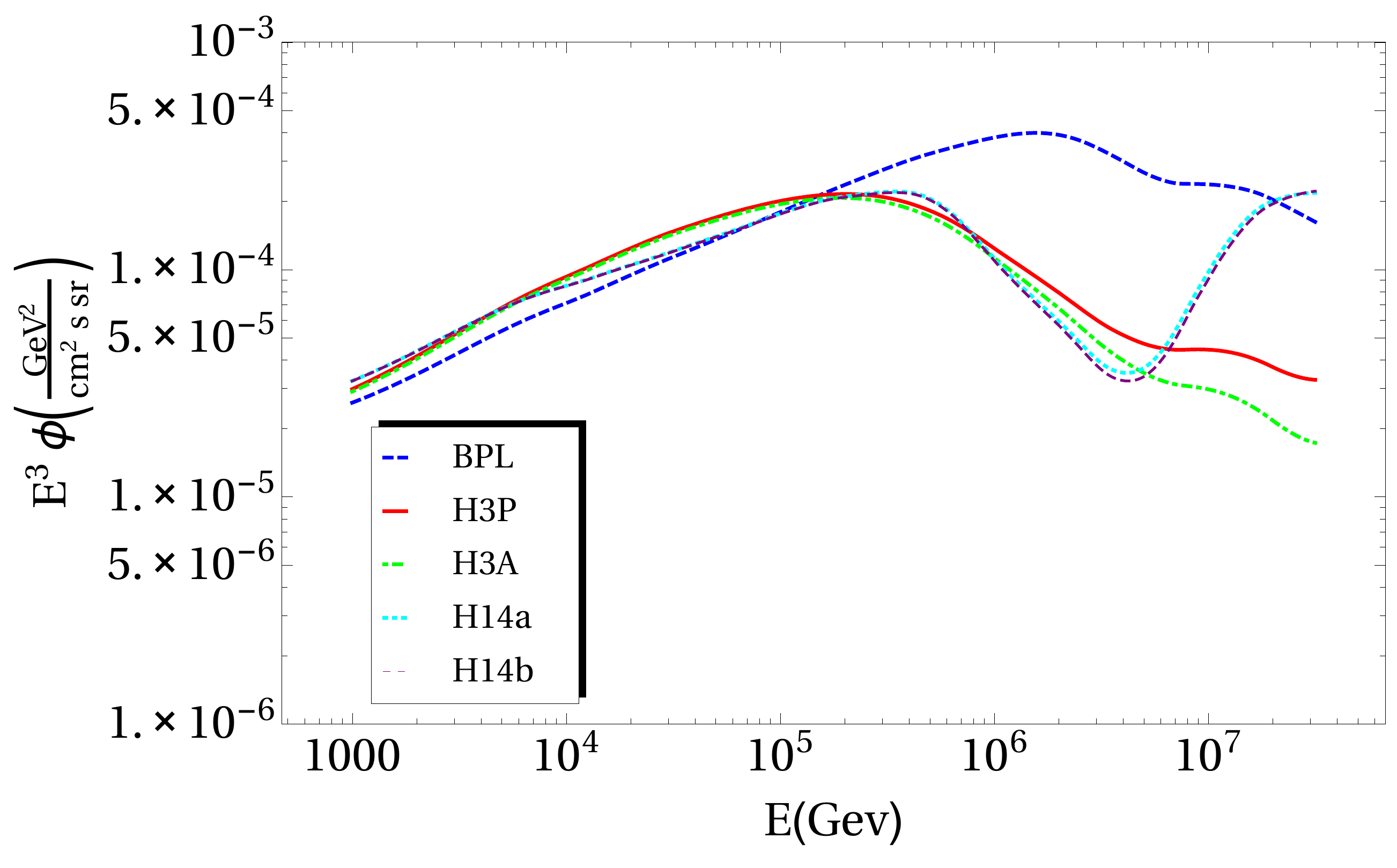}
\includegraphics[scale=0.27]{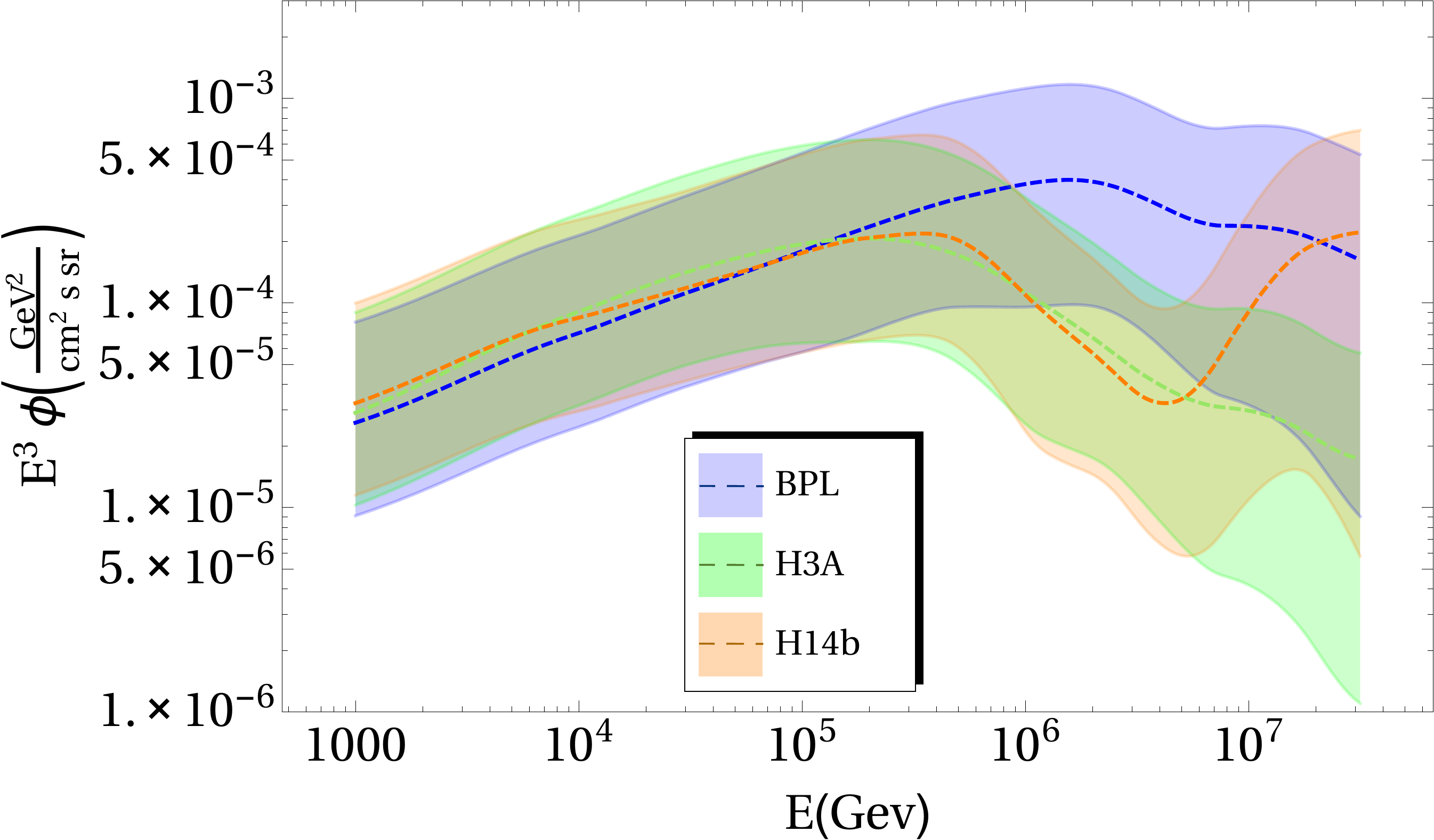}
\caption{\small Left: Comparison of the central values of our
  calculation of the prompt neutrino flux for the 5 different
  parameterisations of the cosmic ray flux.  Right: Comparison of the
  calculations including the theory uncertainty band for the BPL, H3A
  and H14b cosmic ray spectra.}
 \label{fig:centralfluxes}
\end{figure}

In figure~\ref{fig:centralfluxes} we compare our prediction for the
prompt flux for all 5 parameterisations of the cosmic ray flux studied
in this work: BPL, H3P, H3A, H14a and H14b.  For energies
$\lesssim 10^{7}$ GeV, the results for the 4 recent spectra are in
reasonably good agreement with each other but consistently below the
result with the BPL spectrum, with the maximal difference around
$4 \times 10^6$ GeV, where the BPL result is an order of magnitude
larger. At very high neutrino energies $\gtrsim 10^7$ GeV, the recent
H14 parameterisations \cite{Gaisser:2013bla, Stanev:2014mla} yield a
prompt neutrino flux substantially \emph{larger} than with the H3
parameterisations \cite{Gaisser:2011cc}.

In the right plot of figure~\ref{fig:centralfluxes} we perform a
similar comparison, this time between the predictions using the BPL,
H3A and H14b cosmic ray spectra, including in each case the
corresponding theory uncertainty band (which has the same relative
size in all cases, since it arises from the common input of the
$Z_{pm}$ moment).  It is clear that given these uncertainties, the
results for the H3A and H14b parameterisations cannot be
distinguished.
  
Another important input is the choice of PDF set, since knowledge of
the PDF is required at low-$Q^2$ and very small-$x$ where experimental
constraints are generally poor.  In the present calculation, this
uncertainty is substantially reduced by the use of the LHCb charm
production data to constrain the small-$x$ gluon
\cite{Gauld:2015yia}. We now compare our baseline result for the
prompt flux, obtained with the NNPDF3.0+LHCb PDF set (denoted by
NNPDF3.0L), with the central prediction obtained using other PDF
sets:\footnote{Not all available PDF sets can be used for this
  calculation since some of them return negative (unphysical)
  inclusive charm production cross-sections at high-energies, arising
  from a negative gluon at small-$x$.}  ABM11~\cite{Alekhin:2012ig},
CT14~\cite{Dulat:2015mca}, HERAPDF1.5~\cite{H1:2015mha} and
MMHT14~\cite{Harland-Lang:2014zoa}, in all cases at NLO.  For each PDF
set, the {\tt POWHEG} calculation has been set up to include the
required scheme modification terms; for instance, when $n_f=5$ PDFs
are used as input, the scheme transformation terms from $n_f=3$ to
$n_f=5$ are included~\cite{Gauld:2015lxa,Gauld:2015yia}.

\begin{figure}[t]
\centering 
\includegraphics[scale=0.4]{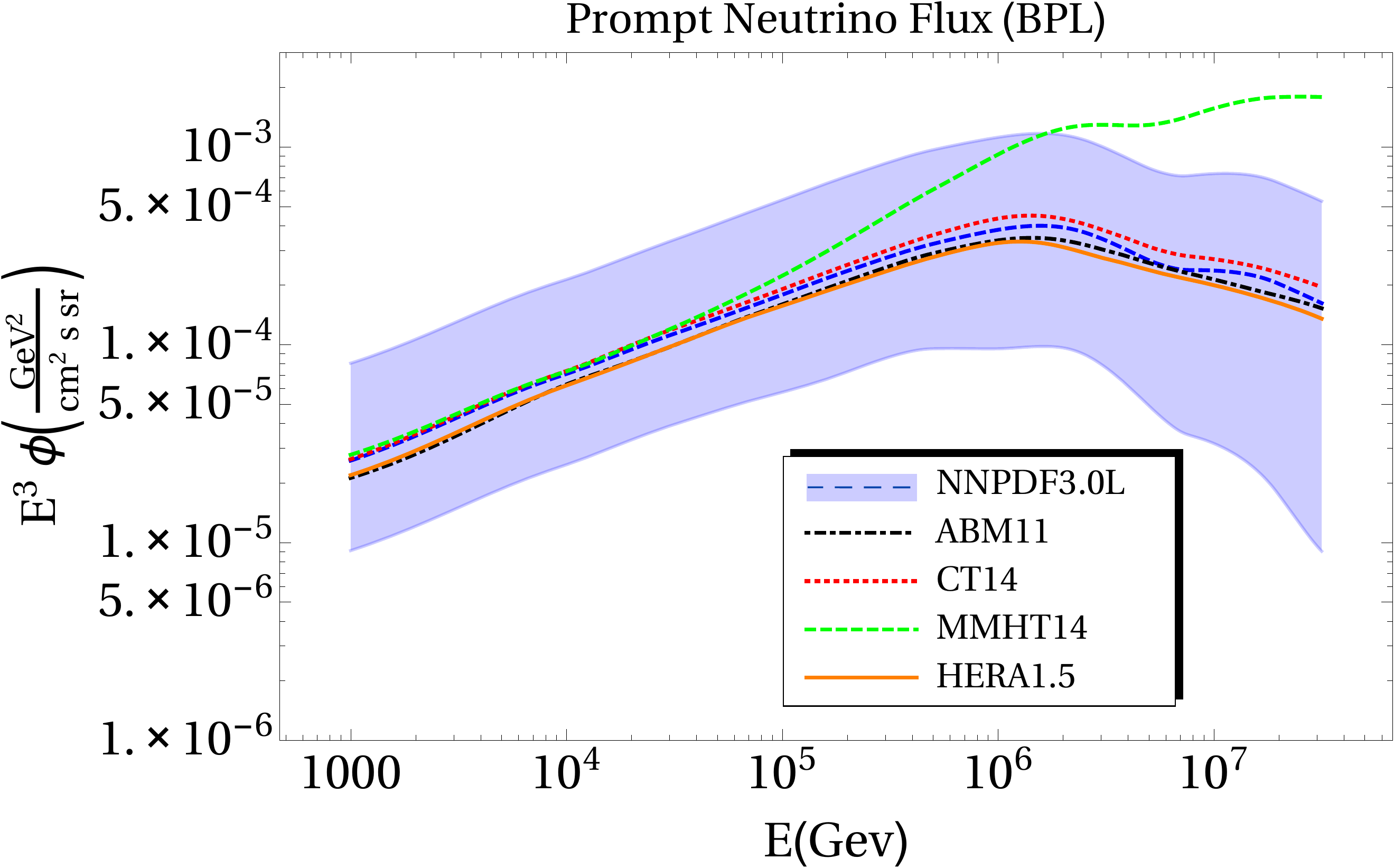}
\caption{\small Comparison of our baseline calculation and its
  uncertainty using the NNPDF3.0L set \cite{Gauld:2015yia}, with the
  corresponding central results using other PDFs as input: ABM11
  \cite{Alekhin:2012ig}, CT14 \cite{Dulat:2015mca}, HERAPDF1.5
  \cite{H1:2015mha} and MMHT14 \cite{Harland-Lang:2014zoa}. All
  calculations assume the BPL cosmic ray spectrum.}
\label{fig:pdfcomp}
\end{figure}

Results for the prompt flux using different PDF sets and the BPL
cosmic ray spectrum are shown in figure~\ref{fig:pdfcomp} where the
total theory uncertainty is shown for NNPDF3.0L only. All PDF sets
yield results in good agreement, except for MMHT14 which yields a
substantially larger flux at energies above $10^5$ GeV.

Thus the choice of PDF set is (with the exception of MMHT14)
\emph{not} important for the central value of the calculated
flux. However it should be emphasised that the theory uncertainty
band, which is shown here only for NNPDF3.0L, would have been much
larger had we not included the LHCb charm hadroproduction data to
reduce the uncertainty in the small-$x$ gluon
\cite{Gauld:2015yia}. Thus our estimate of the uncertainty in the
prompt neutrino flux is more robust than for all other calculations to
date, and accordingly we advocate its use for inferring a \emph{lower}
limit which can be used as a prior in analyses of experimental data.

\subsection{Comparison with previous calculations}

\begin{figure}[t]
\centering 
\includegraphics[scale=0.4]{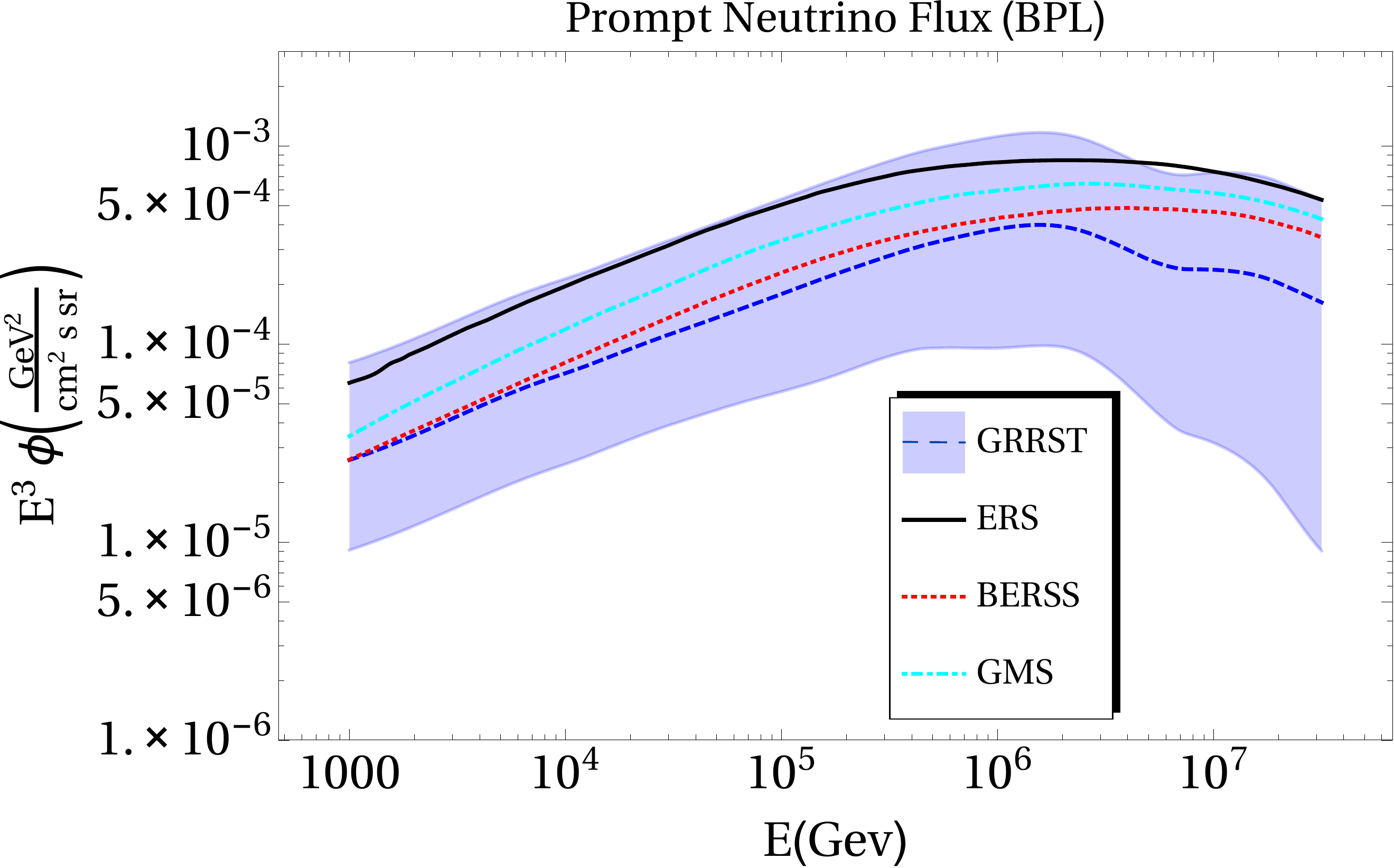}
\caption{\small Comparison of our calculation (GRRST) with the central
  values from ERS \cite{Enberg:2008te}, BERSS
  \cite{Bhattacharya:2015jpa} and GMS \cite{Garzelli:2015psa}, all
  calculated using the BPL cosmic ray spectrum.}
\label{fig:flux2}
\end{figure}

In figure~\ref{fig:flux2} we compare our result with the central values
from the ERS~\cite{Enberg:2008te}, BERSS~\cite{Bhattacharya:2015jpa}
and GMS~\cite{Garzelli:2015psa} calculations, all using the BPL cosmic
ray flux as input.  The relative differences would change only mildly
if a different cosmic ray flux parameterisation was used as input.

The central values of these three previous calculations are contained
within the total theory uncertainty band of our result.  Our central
value is close to BERSS, but systematically smaller than GMS, while
the benchmark ERS result is at the upper end of the theory uncertainty
band.  Note that the BERSS calculation is based on the CT10 NLO PDF
set~\cite{Gao:2013xoa} while the GMS calculation uses the ABM11 PDF
set~\cite{Alekhin:2012ig}, neither of which incorporate the recent
LHCb charm hadroproduction data. The ERS calculation was not based on
pQCD at all, but the empirical `colour dipole model'. It is evident
that there is now some stability in calculations of the prompt
neutrino flux and that in particular a theoretical lower limit can be
set (subject of course to the large systematic uncertainty in the
parameterisation of the incoming cosmic ray flux).

\subsection{Spectral index of the prompt neutrino flux}

\begin{figure}[t]
\centering 
\includegraphics[scale=0.4]{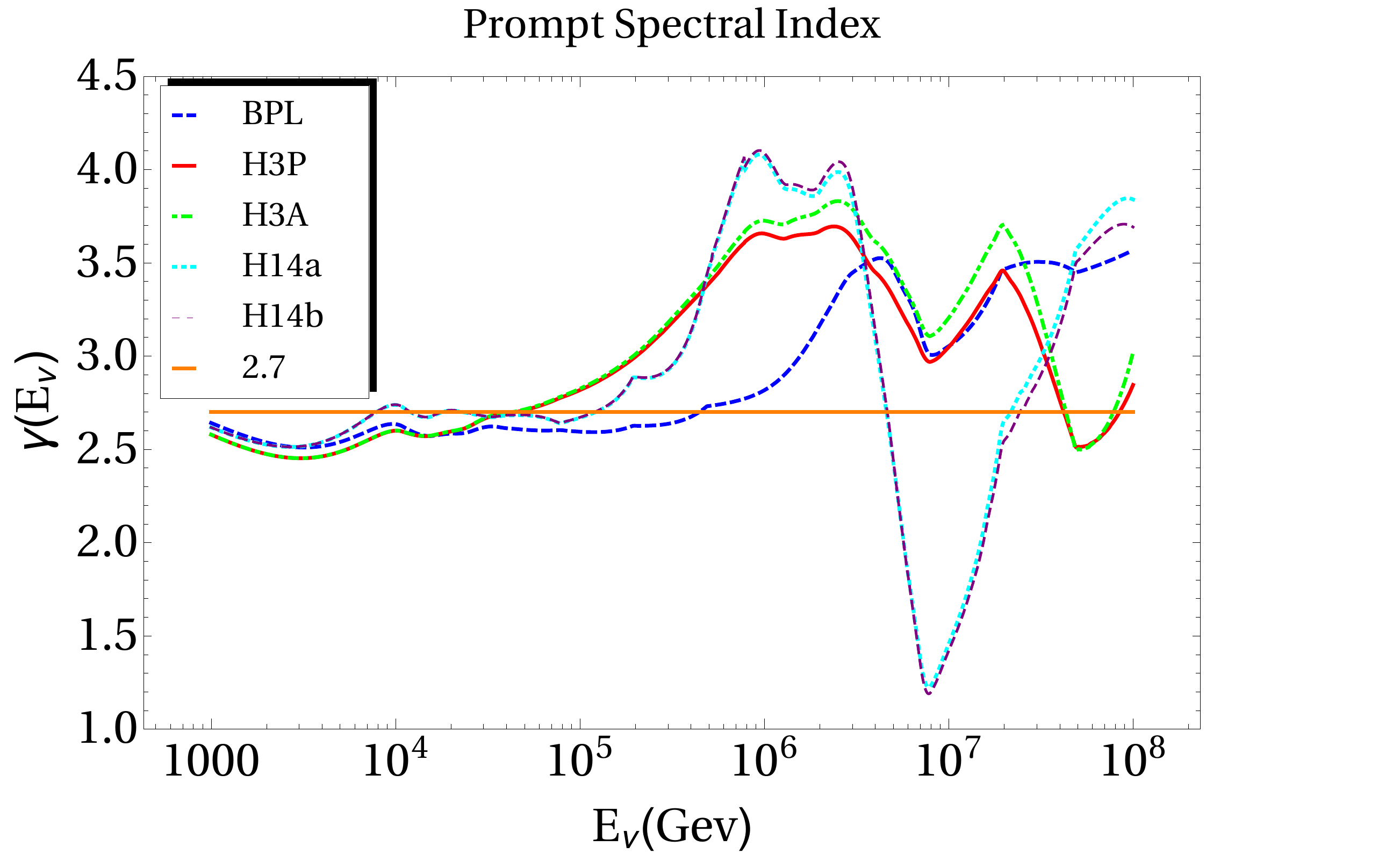}
\caption{\small The effective spectral index $\gamma(E_\nu)$ for the
  central value of our calculation of the prompt neutrino flux for all
  5 parameterisations of the cosmic ray flux.  For reference we show a
  line with $\gamma=2.7$, which is the usual expectation.}
\label{fig:spectralindex}
\end{figure}

It is useful to extract the local spectral index of the prompt
neutrino flux, defined as: 
\begin{equation}
\label{eq:spectralindex}
   \gamma(E_{\nu})\equiv -\frac{d \ln \phi_{\nu}(E_{\nu}) }{d \ln
     E_{\nu}} \, , \quad \text{where} \quad
   \phi_{\nu}(E_{\nu})=A(E_{\nu})\,E_{\nu}^{-\gamma(E_{\nu})} \, , 
\end{equation}
in order to compare with the standard expectation that
$\gamma \simeq 2.7$. Both are shown in figure~\ref{fig:spectralindex}
which illustrates that above $10^5$ GeV the na\"ive scaling is not
obeyed. The BPL, H3P and H3A cosmic ray fluxes all yield a a prompt
neutrino spectrum which falls off more steeply, while with the H14a
and H14b fluxes a harder spectrum is obtained (it is worth keeping in
mind that at very high energies, above $\sim 50$ PeV, charmed mesons
too will begin to lose energy by interaction with air nuclei before
decaying, and at this point the fall-off of the prompt neutrino flux
with $E_{\nu}$ will start to become similar to that of the
conventional flux.).

This indicates that a extraction of the prompt flux from a fit to data
(including both the conventional flux and a cosmic signal) requires
the full calculation of $\phi_{\nu}(E_{\nu})$ as a prior, with the
overall normalisation left free but bounded by the total uncertainty
band shown in figure~\ref{fig:flux1}. At a minimum, the lower limit on
the prompt neutrino flux should be used as a prior, rather than
allowing it to be zero as in current analyses \cite{Schoenen:2015ipa}.

\section{Outlook}
\label{sec:delivery}

We have presented predictions for the flux of prompt neutrinos arising
from the decays of charmed mesons produced in the collisions of high
energy cosmic rays in the atmosphere. Our calculation of charm
production at high energy makes extensive use of NLO Monte Carlo event
generators and parton distribution functions.  The novelty of our
approach is that it has been validated with the 7 TeV charm
cross-sections measured by the LHCb experiment, and found to be
consistent with the more recent 13 TeV measurements.

As input we have used the NNPDF3.0+LHCb PDF set, where the inclusion
of the LHCb 7 TeV data substantially reduces the PDF uncertainties in
the small-$x$ gluon. We include theory uncertainties arising from
PDFs, missing higher-orders, and the value of $m_c$.

We have studied the dependence of our result on the choice of input
cosmic ray fluxes, including the most recent parameterisations, and on
the choice of input PDF set.  Our predictions have been compared with
other calculations, in particular with ERS \cite{Enberg:2008te}, BERSS
\cite{Bhattacharya:2015jpa} and GMS \cite{Garzelli:2015psa}.  All
three calculations are within the uncertainty band of our result,
though our central value is the lowest.  Our result is just consistent
with the current experimental upper limit, suggesting that the prompt
neutrino flux will be detected soon.

Our result for the prompt neutrino flux $\phi_\nu(E_\nu)$ and its
uncertainty, evaluated for 5 different input cosmic ray flux
parameterisations, is available in terms of a {\tt C++} interpolation
code from \url{https://promptnuflux.hepforge.org}.  The interpolation
tables can be used for neutrino energies between $10^3$ and $10^{7.5}$
GeV.

Since our calculations of charm hadroproduction have been validated
with LHCb data, failure to detect the prompt neutrino flux in the
range indicated would imply a flaw in the input assumptions, e.g. the
cosmic ray flux parameterisation or possibly the $Z$-moment approach
itself (e.g. the scalings in
eqs.~\eqref{eq:Zpp}-\eqref{eq:lambdaH}). The latter can only be
addressed by performing a full Monte Carlo simulation as in
\cite{Gondolo:1995fq} with updated QCD tools and data.

\acknowledgments

The authors thank Anna Stasto, Rikard Enberg, Mary Hall Reno and Atri
Bhattacharya for providing the BERSS results, and Maria Garzelli and
Sven Moch for providing the GMS results. S.~S. is grateful to Tom Gaisser,
Francis Halzen, Alexander Kappes and Sebastian Schoenen for helpful
discussions and especially Teresa Montaruli and Anne Schukraft for the
{\tt NeutrinoFlux} code. Rosa Coniglione pointed out an error
in our plotting of the conventional flux in fig.\ref{fig:conventional} 
(c.f. BERSS \cite{Bhattacharya:2015jpa}) and Gary Binder provided the correct
values. R.~G. acknowledges useful comments from Uli Haisch
and Jon Harrison. J.~T. is grateful to Jurgen Rohrwild for useful
discussions.  The work of J.~R. and L.~R. was supported by the ERC
Starting Grant `PDF4BSM' and a STFC Rutherford Fellowship and Grant
awarded to J.~R. (ST/K005227/1 and ST/M003787/1). S.~S. acknowledges a
DNRF Niels Bohr Professorship.  J.~T.  acknowledges support from
Hertford College.

\bibliography{GRRST-fluxes}

\providecommand{\href}[2]{#2}\begingroup\raggedright\begin{thebibliography}{10}

\bibitem{Halzen:2010yj}
F.~Halzen and S.~R. Klein, {\it {IceCube: An instrument for neutrino
  astronomy}},  {\em Rev. Sci. Instrum.} {\bf 81} (2010) 081101,
  [\href{http://arxiv.org/abs/1007.1247}{{\tt arXiv:1007.1247}}].

\bibitem{Aartsen:2013bka}
{\bf IceCube} Collaboration, M.~Aartsen et~al., {\it {First observation of
  PeV-energy neutrinos with IceCube}},  {\em Phys.Rev.Lett.} {\bf 111} (2013)
  021103, [\href{http://arxiv.org/abs/1304.5356}{{\tt arXiv:1304.5356}}].

\bibitem{Aartsen:2013jdh}
{\bf IceCube} Collaboration, M.~Aartsen et~al., {\it {Evidence for high-energy
  extraterrestrial neutrinos at the IceCube detector}},  {\em Science} {\bf
  342} (2013) 1242856, [\href{http://arxiv.org/abs/1311.5238}{{\tt
  arXiv:1311.5238}}].

\bibitem{Aartsen:2014gkd}
{\bf IceCube} Collaboration, M.~Aartsen et~al., {\it {Observation of
  high-energy astrophysical neutrinos in three years of IceCube data}},  {\em
  Phys.Rev.Lett.} {\bf 113} (2014) 101101,
  [\href{http://arxiv.org/abs/1405.5303}{{\tt arXiv:1405.5303}}].

\bibitem{Aartsen:2015ivb}
{\bf IceCube} Collaboration, M.~G. Aartsen et~al., {\it {Flavor Ratio of
  Astrophysical Neutrinos above 35 TeV in IceCube}},  {\em Phys. Rev. Lett.}
  {\bf 114} (2015), no.~17 171102, [\href{http://arxiv.org/abs/1502.03376}{{\tt
  arXiv:1502.03376}}].

\bibitem{Aartsen:2015rwa}
{\bf IceCube} Collaboration, M.~G. Aartsen et~al., {\it {Evidence for
  Astrophysical Muon Neutrinos from the Northern Sky with IceCube}},  {\em
  Phys. Rev. Lett.} {\bf 115} (2015), no.~8 081102,
  [\href{http://arxiv.org/abs/1507.04005}{{\tt arXiv:1507.04005}}].

\bibitem{Aartsen:2015knd}
{\bf IceCube} Collaboration, M.~G. Aartsen et~al., {\it {A combined
  maximum-likelihood analysis of the high-energy astrophysical neutrino flux
  measured with IceCube}},  {\em Astrophys. J.} {\bf 809} (2015), no.~1 98,
  [\href{http://arxiv.org/abs/1507.03991}{{\tt arXiv:1507.03991}}].

\bibitem{Barr:2004br}
G.~Barr, T.~Gaisser, P.~Lipari, S.~Robbins, and T.~Stanev, {\it {A
  three-dimensional calculation of atmospheric neutrinos}},  {\em Phys.Rev.}
  {\bf D70} (2004) 023006, [\href{http://arxiv.org/abs/astro-ph/0403630}{{\tt
  astro-ph/0403630}}].

\bibitem{GonzalezGarcia:2006ay}
M.~C. Gonzalez-Garcia, M.~Maltoni, and J.~Rojo, {\it {Determination of the
  atmospheric neutrino fluxes from atmospheric neutrino data}},  {\em JHEP}
  {\bf 10} (2006) 075, [\href{http://arxiv.org/abs/hep-ph/0607324}{{\tt
  hep-ph/0607324}}].

\bibitem{Honda:2006qj}
M.~Honda, T.~Kajita, K.~Kasahara, S.~Midorikawa, and T.~Sanuki, {\it
  {Calculation of atmospheric neutrino flux using the interaction model
  calibrated with atmospheric muon data}},  {\em Phys.Rev.} {\bf D75} (2007)
  043006, [\href{http://arxiv.org/abs/astro-ph/0611418}{{\tt
  astro-ph/0611418}}].

\bibitem{Honda:2011nf}
M.~Honda, T.~Kajita, K.~Kasahara, and S.~Midorikawa, {\it {Improvement of low
  energy atmospheric neutrino flux calculation using the JAM nuclear
  interaction model}},  {\em Phys. Rev.} {\bf D83} (2011) 123001,
  [\href{http://arxiv.org/abs/1102.2688}{{\tt arXiv:1102.2688}}].

\bibitem{Halzen:2013dva}
F.~Halzen, {\it {The highest energy neutrinos: first evidence for cosmic
  origin}},  {\em Nuovo Cim.} {\bf C037} (2014), no.~03 117--132,
  [\href{http://arxiv.org/abs/1311.6350}{{\tt arXiv:1311.6350}}]. [Astron.
  Nachr.335,507(2014)].

\bibitem{Anchordoqui:2013dnh}
L.~A. Anchordoqui et~al., {\it {Cosmic Neutrino Pevatrons: A Brand New Pathway
  to Astronomy, Astrophysics, and Particle Physics}},  {\em JHEAp} {\bf 1-2}
  (2014) 1--30, [\href{http://arxiv.org/abs/1312.6587}{{\tt arXiv:1312.6587}}].

\bibitem{Lipari:1993hd}
P.~Lipari, {\it {Lepton spectra in the Earth's atmosphere}},  {\em
  Astropart.Phys.} {\bf 1} (1993) 195--227.

\bibitem{Pasquali:1998ji}
L.~Pasquali, M.~Reno, and I.~Sarcevic, {\it {Lepton fluxes from atmospheric
  charm}},  {\em Phys.Rev.} {\bf D59} (1999) 034020,
  [\href{http://arxiv.org/abs/hep-ph/9806428}{{\tt hep-ph/9806428}}].

\bibitem{Enberg:2008te}
R.~Enberg, M.~H. Reno, and I.~Sarcevic, {\it {Prompt neutrino fluxes from
  atmospheric charm}},  {\em Phys.Rev.} {\bf D78} (2008) 043005,
  [\href{http://arxiv.org/abs/0806.0418}{{\tt arXiv:0806.0418}}].

\bibitem{Gondolo:1995fq}
P.~Gondolo, G.~Ingelman, and M.~Thunman, {\it {Charm production and high-energy
  atmospheric muon and neutrino fluxes}},  {\em Astropart.Phys.} {\bf 5} (1996)
  309--332, [\href{http://arxiv.org/abs/hep-ph/9505417}{{\tt hep-ph/9505417}}].

\bibitem{Martin:2003us}
A.~Martin, M.~Ryskin, and A.~Stasto, {\it {Prompt neutrinos from atmospheric
  $c\bar{c}$ and $b \bar{b}$ production and the gluon at very small x}},  {\em
  Acta Phys.Polon.} {\bf B34} (2003) 3273--3304,
  [\href{http://arxiv.org/abs/hep-ph/0302140}{{\tt hep-ph/0302140}}].

\bibitem{Gelmini:1999ve}
G.~Gelmini, P.~Gondolo, and G.~Varieschi, {\it {Prompt atmospheric neutrinos
  and muons: NLO versus LO QCD predictions}},  {\em Phys.Rev.} {\bf D61} (2000)
  036005, [\href{http://arxiv.org/abs/hep-ph/9904457}{{\tt hep-ph/9904457}}].

\bibitem{Bhattacharya:2015jpa}
A.~Bhattacharya, R.~Enberg, M.~H. Reno, I.~Sarcevic, and A.~Stasto, {\it
  {Perturbative charm production and the prompt atmospheric neutrino flux in
  light of RHIC and LHC}},  {\em JHEP} {\bf 06} (2015) 110,
  [\href{http://arxiv.org/abs/1502.01076}{{\tt arXiv:1502.01076}}].

\bibitem{Engel:2015dxa}
F.~Riehn, R.~Engel, A.~Fedynitch, T.~K. Gaisser, and T.~Stanev, {\it {Charm
  production in SIBYLL}},  \href{http://arxiv.org/abs/1502.06353}{{\tt
  arXiv:1502.06353}}.

\bibitem{Fedynitch:2015zma}
A.~Fedynitch, R.~Engel, T.~K. Gaisser, F.~Riehn, and T.~Stanev, {\it {Calculation 
of conventional and prompt lepton fluxes at very high energy}}, 
\href{http://arxiv.org/abs/1503.00544}{{\tt arXiv:1503.00544}}.

\bibitem{Arguelles:2015wba}
C.~A. Arguelles, F.~Halzen, L.~Will, M.~Kroll, and M.~H. Reno, {\it {The
  high-energy behavior of photon, neutrino and proton cross sections}},
  \href{http://arxiv.org/abs/1504.06639}{{\tt arXiv:1504.06639}}.

\bibitem{Garzelli:2015psa}
M.~V. Garzelli, S.~Moch, and G.~Sigl, {\it {Lepton fluxes from atmospheric
  charm revisited}},  {\em JHEP} {\bf 10} (2015) 115,
  [\href{http://arxiv.org/abs/1507.01570}{{\tt arXiv:1507.01570}}].

\bibitem{Aartsen:2014muf}
{\bf IceCube} Collaboration, M.~G. Aartsen et~al., {\it {Atmospheric and
  astrophysical neutrinos above 1 TeV interacting in IceCube}},  {\em Phys.
  Rev.} {\bf D91} (2015), no.~2 022001,
  [\href{http://arxiv.org/abs/1410.1749}{{\tt arXiv:1410.1749}}].

\bibitem{Schoenen:2015ipa}
S.~Schoenen, {\it Talk at the IPA workshop, Madison,
  \url{https://events.icecube.wisc.edu/getFile.py/access?contribId=89&sessionId=42&resId=0&materialId=slides&confId=68}}
  .

\bibitem{Aaij:2013noa}
{\bf LHCb} Collaboration, R.~Aaij et~al., {\it {Measurement of B meson
  production cross-sections in proton-proton collisions at $\sqrt{s}$ = 7
  TeV}},  {\em JHEP} {\bf 1308} (2013) 117,
  [\href{http://arxiv.org/abs/1306.3663}{{\tt arXiv:1306.3663}}].

\bibitem{Aaij:2013mga}
{\bf LHCb} Collaboration, R.~Aaij et~al., {\it {Prompt charm production in pp
  collisions at sqrt(s)=7 TeV}},  {\em Nucl.Phys.} {\bf B871} (2013) 1--20,
  [\href{http://arxiv.org/abs/1302.2864}{{\tt arXiv:1302.2864}}].

\bibitem{Ball:2014uwa}
{\bf NNPDF} Collaboration, R.~D. Ball et~al., {\it {Parton distributions for
  the LHC Run II}},  {\em JHEP} {\bf 1504} (2015) 040,
  [\href{http://arxiv.org/abs/1410.8849}{{\tt arXiv:1410.8849}}].

\bibitem{Zenaiev:2015rfa}
{\bf PROSA} Collaboration, O.~Zenaiev et~al., {\it {Impact of heavy-flavour
  production cross sections measured by the LHCb experiment on parton
  distribution functions at low x}},  {\em Eur. Phys. J.} {\bf C75} (2015),
  no.~8 396, [\href{http://arxiv.org/abs/1503.04581}{{\tt arXiv:1503.04581}}].

\bibitem{Alekhin:2014irh}
S.~Alekhin et~al., {\it {HERAFitter}},  {\em Eur. Phys. J.} {\bf C75} (2015),
  no.~7 304, [\href{http://arxiv.org/abs/1410.4412}{{\tt arXiv:1410.4412}}].

\bibitem{Cacciari:2001td}
M.~Cacciari, S.~Frixione, and P.~Nason, {\it {The p(T) spectrum in heavy
  flavour photoproduction}},  {\em JHEP} {\bf 0103} (2001) 006,
  [\href{http://arxiv.org/abs/hep-ph/0102134}{{\tt hep-ph/0102134}}].

\bibitem{Nason:2004rx}
P.~Nason, {\it {A new method for combining NLO QCD with shower Monte Carlo
  algorithms}},  {\em JHEP} {\bf 0411} (2004) 040,
  [\href{http://arxiv.org/abs/hep-ph/0409146}{{\tt hep-ph/0409146}}].

\bibitem{Frixione:2007vw}
S.~Frixione, P.~Nason, and C.~Oleari, {\it {Matching NLO QCD computations with
  parton shower simulations: the POWHEG method}},  {\em JHEP} {\bf 0711} (2007)
  070, [\href{http://arxiv.org/abs/0709.2092}{{\tt arXiv:0709.2092}}].

\bibitem{Alioli:2010xd}
S.~Alioli, P.~Nason, C.~Oleari, and E.~Re, {\it {A general framework for
  implementing NLO calculations in shower Monte Carlo programs: the POWHEG
  BOX}},  {\em JHEP} {\bf 1006} (2010) 043,
  [\href{http://arxiv.org/abs/1002.2581}{{\tt arXiv:1002.2581}}].

\bibitem{Alwall:2014hca}
J.~Alwall, R.~Frederix, S.~Frixione, V.~Hirschi, F.~Maltoni, et~al., {\it {The
  automated computation of tree-level and next-to-leading order differential
  cross sections, and their matching to parton shower simulations}},  {\em
  JHEP} {\bf 1407} (2014) 079, [\href{http://arxiv.org/abs/1405.0301}{{\tt
  arXiv:1405.0301}}].

\bibitem{Aaij:2015bpa}
{\bf LHCb} Collaboration, R.~Aaij et~al., {\it {Measurements of prompt charm
  production cross-sections in $pp$ collisions at $\sqrt{s}$ = 13 TeV}},
  \href{http://arxiv.org/abs/1510.01707}{{\tt arXiv:1510.01707}}.

\bibitem{Gaisser:2013bla}
T.~K. Gaisser, T.~Stanev, and S.~Tilav, {\it {Cosmic Ray Energy Spectrum from
  Measurements of Air Showers}},  {\em Front.Phys.China} {\bf 8} (2013)
  748--758, [\href{http://arxiv.org/abs/1303.3565}{{\tt arXiv:1303.3565}}].

\bibitem{Stanev:2014mla}
T.~Stanev, T.~K. Gaisser, and S.~Tilav, {\it {High energy cosmic rays: sources
  and fluxes}},  {\em Nucl.Instrum.Meth.} {\bf A742} (2014) 42--46.

\bibitem{Seo:2012pw}
E.~S. Seo, {\it {Direct measurements of cosmic rays using balloon borne
  experiments}},  {\em Astropart. Phys.} {\bf 39-40} (2012) 76--87.

\bibitem{Kampert:2012mx}
K.-H. Kampert and M.~Unger, {\it {Measurements of the cosmic ray composition
  with air shower experiments}},  {\em Astropart. Phys.} {\bf 35} (2012)
  660--678, [\href{http://arxiv.org/abs/1201.0018}{{\tt arXiv:1201.0018}}].

\bibitem{Peixoto:2015ava}
C.~J. Todero~Peixoto, V.~de~Souza, and P.~L. Biermann, {\it {Cosmic rays: the
  spectrum and chemical composition from $10^{10}$ to $10^{20}$ eV}},  {\em
  JCAP} {\bf 1507} (2015), no.~07 042,
  [\href{http://arxiv.org/abs/1502.00305}{{\tt arXiv:1502.00305}}].

\bibitem{Gaisser:2011cc}
T.~K. Gaisser, {\it {Spectrum of cosmic-ray nucleons, kaon production, and the
  atmospheric muon charge ratio}},  {\em Astropart. Phys.} {\bf 35} (2012)
  801--806, [\href{http://arxiv.org/abs/1111.6675}{{\tt arXiv:1111.6675}}].

\bibitem{Hillas:2006ms}
A.~M. Hillas, {\it {Cosmic rays: Recent progress and some current questions}},
  in {\em {Conference on Cosmology, Galaxy Formation and Astro-Particle Physics
  on the Pathway to the SKA Oxford, England, April 10-12, 2006}}, 2006.
\newblock \href{http://arxiv.org/abs/astro-ph/0607109}{{\tt astro-ph/0607109}}.

\bibitem{Gaisser:1990vg}
T.~K. Gaisser, {\em {Cosmic rays and particle physics}}.
\newblock Cambridge University Press, 1990.

\bibitem{Kalmykov:1993qe}
N.~N. Kalmykov and S.~S. Ostapchenko, {\it {The nucleus-nucleus interaction,
  nuclear fragmentation, and fluctuations of extensive air showers}},  {\em
  Phys. Atom. Nucl.} {\bf 56} (1993) 346--353. [Yad. Fiz.56N3,105(1993)].

\bibitem{Mielke:1994un}
H.~H. Mielke, M.~Foeller, J.~Engler, and J.~Knapp, {\it {Cosmic ray hadron flux
  at sea level up to 15 TeV}},  {\em J. Phys.} {\bf G20} (1994) 637--649.

\bibitem{Bugaev:1998bi}
E.~Bugaev, A.~Misaki, V.~A. Naumov, T.~Sinegovskaya, S.~Sinegovsky, et~al.,
  {\it {Atmospheric muon flux at sea level, underground and underwater}},  {\em
  Phys.Rev.} {\bf D58} (1998) 054001,
  [\href{http://arxiv.org/abs/hep-ph/9803488}{{\tt hep-ph/9803488}}].

\bibitem{Heck:1998vt}
D.~Heck, G.~Schatz, T.~Thouw, J.~Knapp, and J.~N. Capdevielle, {\it {CORSIKA: A
  Monte Carlo code to simulate extensive air showers}}, .

\bibitem{Ahn:2009wx}
E.-J. Ahn, R.~Engel, T.~K. Gaisser, P.~Lipari, and T.~Stanev, {\it {Cosmic ray
  interaction event generator SIBYLL 2.1}},  {\em Phys. Rev.} {\bf D80} (2009)
  094003, [\href{http://arxiv.org/abs/0906.4113}{{\tt arXiv:0906.4113}}].

\bibitem{Antchev:2011vs}
G.~Antchev et~al., {\it {First measurement of the total proton-proton cross
  section at the LHC energy of $\sqrt{s}$ =7 TeV}},  {\em Europhys. Lett.} {\bf
  96} (2011) 21002, [\href{http://arxiv.org/abs/1110.1395}{{\tt
  arXiv:1110.1395}}].

\bibitem{Collaboration:2012wt}
{\bf Pierre Auger} Collaboration, P.~Abreu et~al., {\it {Measurement of the
  proton-air cross-section at $\sqrt{s}=57$ TeV with the Pierre Auger
  Observatory}},  {\em Phys. Rev. Lett.} {\bf 109} (2012) 062002,
  [\href{http://arxiv.org/abs/1208.1520}{{\tt arXiv:1208.1520}}].

\bibitem{Sjostrand:2014zea}
T.~Sjöstrand, S.~Ask, J.~R. Christiansen, R.~Corke, N.~Desai, et~al., {\it {An
  introduction to PYTHIA 8.2}},  {\em Comput.Phys.Commun.} {\bf 191} (2015)
  159--177, [\href{http://arxiv.org/abs/1410.3012}{{\tt arXiv:1410.3012}}].

\bibitem{Agashe:2014kda}
{\bf Particle Data Group} Collaboration, K.~A. Olive et~al., {\it {Review of
  Particle Physics}},  {\em Chin. Phys.} {\bf C38} (2014) 090001.

\bibitem{Gauld:2015lxa}
R.~Gauld, {\it {Forward $D$ predictions for $p\rm Pb$ collisions, and
  sensitivity to cold nuclear matter effects}},
  \href{http://arxiv.org/abs/1508.07629}{{\tt arXiv:1508.07629}}.

\bibitem{Khachatryan:2015uja}
{\bf CMS} Collaboration, V.~Khachatryan et~al., {\it {Study of B Meson
  Production in pPb Collisions at $\sqrt{s_{\rm{NN}}}=$=5.02 TeV}},
  \href{http://arxiv.org/abs/1508.06678}{{\tt arXiv:1508.06678}}.

\bibitem{Gauld:2015yia}
R.~Gauld, J.~Rojo, L.~Rottoli, and J.~Talbert, {\it {Charm production in the
  forward region: constraints on the small-x gluon and backgrounds for neutrino
  astronomy}},  {\em JHEP} {\bf 11} (2015) 009,
  [\href{http://arxiv.org/abs/1506.08025}{{\tt arXiv:1506.08025}}].

\bibitem{Skands:2014pea}
P.~Skands, S.~Carrazza, and J.~Rojo, {\it {Tuning PYTHIA 8.1: the Monash 2013
  tune}},  {\em European Physical Journal} {\bf 74} (2014) 3024,
  [\href{http://arxiv.org/abs/1404.5630}{{\tt arXiv:1404.5630}}].

\bibitem{Cacciari:2005uk}
M.~Cacciari, P.~Nason, and C.~Oleari, {\it {A study of heavy flavoured meson
  fragmentation functions in e+ e- annihilation}},  {\em JHEP} {\bf 0604}
  (2006) 006, [\href{http://arxiv.org/abs/hep-ph/0510032}{{\tt
  hep-ph/0510032}}].

\bibitem{Czakon:2015owf}
M.~Czakon, D.~Heymes, and A.~Mitov, {\it {High-precision differential
  predictions for top-quark pairs at the LHC}},
  \href{http://arxiv.org/abs/1511.00549}{{\tt arXiv:1511.00549}}.

\bibitem{Aartsen:2015xup}
{\bf IceCube} Collaboration, M.~G. Aartsen et~al., {\it {Measurement of the Atmospheric $\nu_e$
Spectrum with IceCube}}, {\em Phys.Rev.} {\bf D91} (2015) 122004,
  [\href{http://arxiv.org/abs/1504.03753}{{\tt arXiv:1504.03753}}].

\bibitem{Alekhin:2012ig}
S.~Alekhin, J.~Bl{\"u}mlein, and S.~Moch, {\it {Parton distribution functions
  and benchmark cross sections at NNLO}},  {\em Phys.Rev.} {\bf D86} (2012)
  054009, [\href{http://arxiv.org/abs/1202.2281}{{\tt arXiv:1202.2281}}].

\bibitem{Dulat:2015mca}
S.~Dulat, T.~J. Hou, J.~Gao, M.~Guzzi, J.~Huston, P.~Nadolsky, J.~Pumplin,
  C.~Schmidt, D.~Stump, and C.~P. Yuan, {\it {The CT14 global analysis of
  quantum chromodynamics}},  \href{http://arxiv.org/abs/1506.07443}{{\tt
  arXiv:1506.07443}}.

\bibitem{H1:2015mha}
{\bf ZEUS and H1} Collaboration, S.~Schmidtt, {\it {Combination of measurements
  of inclusive deep inelastic $e^{\pm}p$ scattering cross sections and QCD
  analysis of HERA data}},  \href{http://arxiv.org/abs/1506.06042}{{\tt
  arXiv:1506.06042}}.

\bibitem{Harland-Lang:2014zoa}
L.~Harland-Lang, A.~Martin, P.~Motylinski, and R.~Thorne, {\it {Parton
  distributions in the LHC era: MMHT 2014 PDFs}},  {\em Eur.Phys.J.} {\bf C75}
  (2015), no.~5 204, [\href{http://arxiv.org/abs/1412.3989}{{\tt
  arXiv:1412.3989}}].

\bibitem{Gao:2013xoa}
J.~Gao, M.~Guzzi, J.~Huston, H.-L. Lai, Z.~Li, P.~Nadolsky, J.~Pumplin,
  D.~Stump, and C.~P. Yuan, {\it {CT10 next-to-next-to-leading order global
  analysis of QCD}},  {\em Phys. Rev.} {\bf D89} (2014), no.~3 033009,
  [\href{http://arxiv.org/abs/1302.6246}{{\tt arXiv:1302.6246}}].

\end{thebibliography}\endgroup

\end{document}